\documentclass{sig-alternate-2013}
\usepackage[utf8x]{inputenc}
\usepackage{amsmath,amsfonts}
\usepackage{balance}
\usepackage[pagebackref=true,breaklinks=true,colorlinks=true]{hyperref}
\usepackage[english]{babel}
\usepackage{algorithmic,algorithm}
\usepackage{booktabs}
\usepackage{graphicx}
\usepackage{bbm}
\usepackage{caption}
\usepackage{subcaption}
\newcommand{\neigh}[1]{\mathcal{N}(#1)}
\newcommand{\defeq}{\equiv}
\newcommand{\reals}{\mathbf{R}}
\DeclareMathOperator*{\argmax}{argmax}
\newtheorem{proposition}{Proposition}
\newtheorem{lemma}{Lemma}

\begin{document}

\permission{Copyright is held by the International World Wide Web Conference Committee (IW3C2). IW3C2 reserves the right to provide a hyperlink to the author's site if the Material is used in electronic media.}
\conferenceinfo{WWW 2015,}{May 18--22, 2015, Florence, Italy.}
\copyrightetc{ACM \the\acmcopyr}
\crdata{978-1-4503-3469-3/15/05. \\
    http://dx.doi.org/10.1145/2736277.2741127}

\clubpenalty=10000
\widowpenalty = 10000

\title{Scalable Methods for Adaptively Seeding a Social Network}
%Paradox for Influence Maximization}

\numberofauthors{2}
\author{
\alignauthor
Thibaut Horel\\
       \affaddr{Harvard University}\\
       \email{thorel@seas.harvard.edu}
\alignauthor
Yaron Singer\\
       \affaddr{Harvard University}\\
       \email{yaron@seas.harvard.edu}
}
\date{}

\maketitle
\begin{abstract}
    In recent years, social networking platforms have developed into extraordinary channels for spreading and consuming information. Along with the rise of such infrastructure, there is continuous progress on techniques for spreading information effectively through influential users. In many applications, one is restricted to select influencers from a set of users who engaged with the topic being promoted, and due to the structure of social networks, these users often rank low in terms of their influence potential. An alternative approach one can consider is an adaptive method which selects users in a manner which targets their influential neighbors. The advantage of such an approach is that it leverages the friendship paradox in social networks: while users are often not influential, they often know someone who is.  

Despite the various complexities in such optimization problems, we show that scalable adaptive seeding is achievable. In particular, we develop algorithms for linear influence models with provable approximation guarantees that can be gracefully parallelized. To show the effectiveness of our methods we collected data from various verticals social network users follow. For each vertical, we collected data on the users who responded to a certain post as well as their neighbors, and applied our methods on this data. Our experiments show that adaptive seeding is  scalable, and importantly, that it obtains dramatic improvements over standard approaches of information dissemination.

\end{abstract}

%\terms{Theory, Algorithms, Performance}
%\category{H.4}{Information Systems Applications}{Miscellaneous}
%\category{D.2.8}{Software Engineering}{Metrics}[complexity measures, performance measures]
%\terms{Theory}
%\keywords{ACM proceedings, \LaTeX, text tagging}

\newpage
\section{Introduction}

%The massive adoption of social networking services in recent years creates a unique platform for disseminating information and promoting ideas.  Effective means to spread information through this platform are continuously being developed, and in particular methods for selecting influential users who can trigger large word-of-mouth cascade have been heavily studied throughout the past decade.  Initially studied by Domingos and Richarson~\cite{} and formalized by Kempe  
%
%
%Effective means to spread information through this platform are continuously being developed, and in particular methods for s
The massive adoption of social networking services in recent years creates a unique platform for promoting ideas and spreading information.  Communication through online social networks leaves traces of behavioral data which allow observing, predicting and even engineering processes of information diffusion.  First posed by Domingos and Richardson~\cite{DR01,RD02} and elegantly formulated and further developed by Kempe, Kleinberg, and Tardos~\cite{KKT03}, \emph{influence maximization} is the algorithmic challenge of selecting a fixed number of individuals who can serve as early adopters of a new idea, product, or technology in a manner that will trigger a large cascade in the social network.  
%Since its inception, numerous techniques and improvements have been developed ranging from sophisticated predictive models of influence ~\cite{LAH06,RLK10,BHMW11,ACKS13,manuel2013icml,du13nips} to fast approximation methods~\cite{LKGFVG07,MR07,C08,KDD11,borgs2012influence}. 
%While there has been a great deal of progress on efficient algorithmic methods for this problem and impressive methods for learning models of influence from data, a fundamental problem has been largely overlooked.  

In many cases where influence maximization methods are applied one cannot
select any user in the network but is limited to some subset of users.  As an
example, consider an online retailer who wishes to promote a product through
word-of-mouth by rewarding influential customers who purchased the product.
The retailer is then limited to select influential users from the set of users
who purchased the product. In general, we will call the \emph{core set} the set
of users an influence maximization campaign can access.  When the goal is to
select influential users from the core set, the laws that govern social
networks can lead to poor outcomes.  Due to the heavy-tailed degree
distribution of social networks, high degree nodes are rare, and since
influence maximization techniques often depend on the ability to select high
%(not necessarily the highest)
degree nodes, a naive application of influence
maximization techniques to the core set can become ineffective.\newline

%For a concrete example, consider a scenario where the goal is to select influential users who visit an online store and reward them for promoting a product through their social network.  In such a case, if the users who visit the online store are not influential, even the best techniques for identifying influential users would have poor performance.  

\begin{figure}
    \centering
    \includegraphics[scale=0.55]{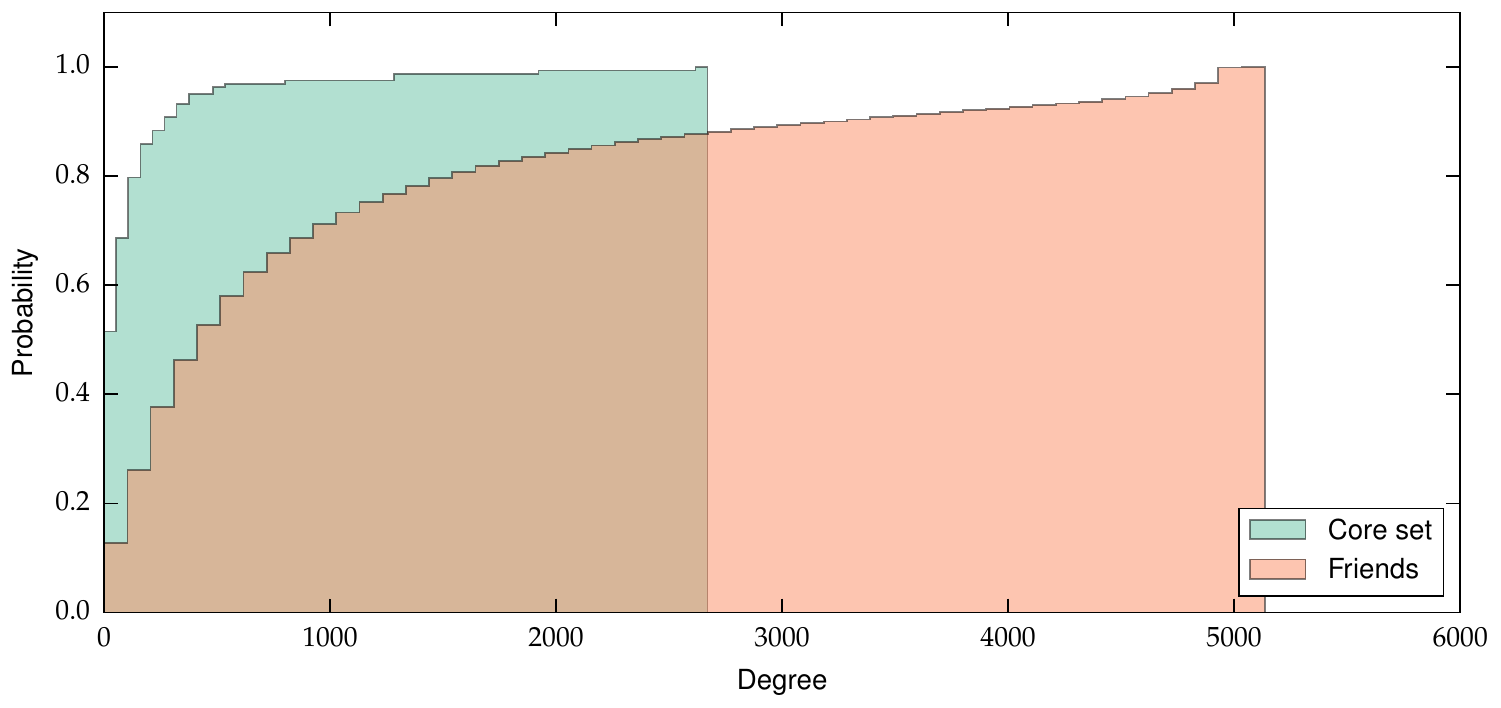}
    \vspace{-15pt}
    \caption{CDF of the degree distribution of users who liked a post by Kiva on Facebook and that of their friends.}
    \label{fig:para}
    \vspace{-10pt}
\end{figure}

\noindent \textbf{An adaptive approach.}
An alternative approach to spending the entire budget on the core set is an
adaptive two-stage approach.  In the first stage, one can spend a fraction of
the budget on the core users so that they invite their friends to participate
in the campaign, then in the second stage spend the rest of the budget
on the influential friends who hopefully have arrived.  The idea behind this
approach is to leverage a structural phenomenon in social networks known as
the friendship paradox~\cite{feld1991}.  Intuitively, the friendship paradox
says that individuals are not likely to have many friends, but they likely
have a friend that does (``your friends have more friends than you'').  In
Figure~\ref{fig:para} we give an example of such an effect by plotting a CDF of
the degree distribution of a core set of users who responded to a post on
Facebook and the degree distribution of their friends.  Remarkably, there are
also formal guarantees of such effects.  Recent work shows that for any network
that has a power law degree distribution and a small fraction of random edges,
there is an asymptotic gap between the average degree of small samples of nodes
and that of their neighbors, with constant probability~\cite{LS13}.  The
implication is that when considering the core users (e.g. those who visit the
online store) as random samples from a social network, any algorithm which can
use their neighbors as influencers will have dramatic improvement over the
direct application of influence maximization.\newline

\noindent \textbf{Warmup.}  Suppose we are given a network, a random set of
core users $X$ and a budget $k$, and the goal is to select a subset of nodes in
$X$ of size $t\leq k$ which has the most influential set of neighbors of size
$k-t$.  For simplicity, assume for now that the influence of a set is simply
its average degree.  If we take the $k/2$ highest degree neighbors of $X$, then
surely there is a set $S$ of size at most $k/2$ in $X$ connected to this set,
and selecting $S$ would be a two-approximation to this problem.  In comparison,
the standard approach of influence maximization is to select the $k$ highest
degree nodes in $X$.  Thus, standard influence maximization would have $k$ of
the most influential nodes in $X$ while the approximation algorithm we propose
has $k/2$ of the most influential nodes from its set of neighbors.  How much
better or worse is it to use this approach over the standard one?  If
the network has a power-law degree distribution with a small fraction of random
edges, and influence is measured in terms of sum of degrees of a set, then the
results of~\cite{LS13} discussed above imply that the two-stage approach
which allows seeding neighbors can do asymptotically (in the size of the
network) better.
%\footnote{\tiny{In fact, the results of~\cite{LS13} hold not only for measures of influence such as degree, but also for measures like coverage functions, and the voter model which we discuss in this paper.}}.  
Thus, at least intuitively, it looks as if two-stage approaches may be worth
investigating.   
\newline

In this paper, our goal is to study the potential benefits of adaptive approaches for influence maximization.  We are largely motivated by the following question.

\begin{center}
\textit{Can adaptive optimization lead to significant improvements in influence maximization?}
\end{center}

To study this question we use the adaptive seeding model recently formalized
in~\cite{singer}.  The main distinctions from the caricature model in the
warmup problem above is that in adaptive seeding the core set $X$ can be
arbitrary (it does not have to be random), and every neighbor of $X$ is assumed
to arrive with some independent probability. These probabilities are used to
model the uncertainty we have in that the neighbors would be interested in
promoting the product, as they are not in the core set.  The goal in adaptive
seeding is to select a subset of nodes in $X$ such that, in expectation over
all possible arrivals of its neighbors, one can select a maximally influential
set of neighbors with the remaining budget.\footnote{\tiny{The model can be
        extended to the case where nodes take on different costs, and results
        we present here largely generalize to such settings as well.  Although
        it seems quite plausible that the probability of attracting neighbors
        could depend on the rewards they receive, the model deliberately
        assumes unit costs, consistent with the celebrated
Kempe-Kleinberg-Tardos model~\cite{KKT03}. Of course, if the likelihood of
becoming an early adopter is inversely proportional to one's influence, then
any influence maximization model loses substance.}}  It is worth noting that
using $X$ to be the entire set of nodes in the network we get the
Kempe-Kleinberg-Tardos model~\cite{KKT03}, and thus adaptive seeding can be
seen as a generalization of this model.\newline

\noindent \textbf{Scalability.}  One of the challenges in adaptive seeding is
scalability.  This is largely due to the stochastic nature of the problem
derived from uncertainty about arrival of neighbors.  The main result
in~\cite{singer} is a constant factor approximation algorithm for well-studied
influence models such as independent cascade and linear threshold which is, at
large, a theoretical triumph.  These algorithms rely on various forms of
sampling, which lead to a significant blowup in the input size.  While such
techniques provide strong theoretical guarantees, for social network data sets
which are often either large or massive, such approaches are inapplicable.  The
main technical challenge we address in this work is how to design scalable
adaptive optimization techniques for influence maximization which do not
require sampling.\newline

\noindent \textbf{Beyond random users.}  The motivation for the adaptive
approach hinges on the friendship paradox, but what if the core set is not
a random sample?  The results in~\cite{LS13} hold when the core set of users is
random but since users who follow a particular topic are not a random sample of
the network, we must somehow evaluate adaptive seeding on representative data
sets.  The experimental challenge is to estimate the prominence of high degree
neighbors in settings that are typical of viral marketing campaigns.
Figure~\ref{fig:para} is a foreshadowing of the experimental methods we used to
show that an effect similar to the friendship paradox exists in such cases as
well.  \newline

\noindent \textbf{Main results.}
Our main results in this paper show that adaptive seeding is a scalable
approach which can dramatically improve upon standard approaches of influence
maximization.  We present a general method that enables designing adaptive
seeding algorithms in a manner that avoids sampling, and thus makes adaptive
seeding scalable to large size graphs.  We use this approach as a basis for
designing two algorithms, both achieving an approximation ratio of $(1-1/e)$
for the adaptive problem.  The first algorithm is implemented through a linear
program, which proves to be extremely efficient over instances where there is
a large budget.  The second approach is a combinatorial algorithm with the same
approximation guarantee which can be easily parallelized, has good theoretical
guarantees on its running time and does well on instances with smaller budgets.
The guarantees of our algorithms hold for linear models of influence,
\emph{i.e.} models for which the influence of a set can be expressed as the sum
of the influence of its members.  While this class does not include models such
as the independent cascade and the linear threshold model, it includes the
well-studied \emph{voter model}~\cite{holley1975ergodic,even-dar} and measures
such as node degree, click-through-rate or retweet measures of users which
serve as natural proxies of influence in many settings~\cite{ZHGS10}.  In
comparison to submodular influence functions, the relative simplicity of linear
models allows making substantial progress on this challenging problem.

We then use these algorithms to conduct a series of experiments to show the
potential of adaptive approaches for influence maximization both on synthetic
and real social networks. The main component of the experiments involved
collecting publicly available data from Facebook on users who expressed
interest (``liked'') a certain post from a topic they follow and data on their
friends.  The premise here is that such users mimic potential participants in
a viral marketing campaign.  The results on these data sets suggest that
adaptive seeding can have dramatic improvements over standard influence
maximization methods.\newline

\noindent \textbf{Paper organization.}
We begin by formally describing the model and the assumptions we make in the
following section.  In Section~\ref{sec:adaptivity} we describe the reduction
of the adaptive seeding problem to a non-adaptive relaxation.  In
Section~\ref{sec:algorithms} we describe our non-adaptive algorithms for
adaptive seeding. In Section~\ref{sec:experiments} we describe our
experiments, and conclude with a brief discussion on related work.
 
%\subsection{Stochastic optimization sans sampling}
%A common approach in stochastic optimization is Sample Average Approximation (SAA).
%The main idea behind SAA is to sample realizations of the second stage, solve the optimization problem on the sampled instances, and average the solution.  Often, when the number of samples is polynomial in the input size of the problem.  In our case the problem is too large.
%
%In this paper we show a new technique for solving stochastic optimization problems which does not use sampling.
%
%
%\subsection{Roadmap}
%The main framework we apply in this paper is to develop non-adaptive algorithms which we will then use to create adaptive solutions. At a high level, the main idea would be to use a particular version of non-adaptive algorithms whose optimal solution is an upper bound to the optimal adaptive policy.  We will then argue that a solution to the non-adaptive problem can be turned into a solution to the adaptive policy, without losing almost any value.  This will therefore reduce our problem to that of designing solutions to the non-adaptive problem we define, for which we develop specialized algorithms.   
%
%Beyond the approximation guarantees, the main advantage in the non-adaptive framework is that unlike standard approaches in stochastic optimization, it avoids using sampling.  As we will later discuss, this dramatically reduces the running time the algorithms, both in theory and practice.  
%
%

\section{Model}
\label{sec:model}
%\subsection{Problem and notations}
%Let us begin by introducing the following notation.
%We will use the following notation to formally discuss the model.  
Given a graph $G=(V,E)$, for a node $v\in V$ we denote by $\neigh{v}$
the neighborhood of $v$. By extension, for any subset of nodes $S\subseteq V$,
$\neigh{S}\defeq \bigcup_{v\in S}\neigh{v}$ will denote the neighborhood of
$S$. The notion of influence in the graph is captured by a function
$f:2^{|V|}\rightarrow \reals_+$ mapping a subset of nodes to a non-negative
influence value.\newline

\noindent \textbf{The adaptive seeding model.} 
The input of the \emph{adaptive seeding} problem is a \emph{core set} of nodes
$X\subseteq V$ and for any node $u\in\neigh{X}$ a probability $p_u$ that $u$
realizes if one of its neighbor in $X$ is seeded. We will write $m=|X|$ and
$n=|\neigh{X}|$ the parameters controlling the input size. The seeding process
is the following:
\begin{enumerate}
    \item \emph{Seeding:} the seeder selects a subset of nodes $S\subseteq
    X$ in the core set.
    \item \emph{Realization of the neighbors:} every node $u\in\neigh{S}$
    realizes independently with probability $p_u$. We denote by
    $R\subseteq\neigh{S}$ the subset of nodes that is realized during this
    stage.
    \item \emph{Influence maximization:} the seeder selects the set of nodes
    $T\subseteq R$ that maximizes the influence function $f$.
\end{enumerate}

There is a budget constraint $k$ on the total number of nodes that can be
selected: $S$ and $T$ must satisfy $|S|+|T|\leq k$. The seeder chooses the set
$S$ before observing the realization $R$ and thus wishes to select optimally in
expectation over all such possible realizations. Formally, the objective can be
stated as:
\begin{equation}\label{eq:problem}
    \begin{split}
        &\max_{S\subseteq X} \sum_{R\subseteq\neigh{S}} p_R
        \max_{\substack{T\subseteq R\\|T|\leq k-|S|}}f(T)\\
        &\text{s.t. }|S|\leq k
    \end{split}
\end{equation}
where $p_R$ is the probability that the set $R$ realizes,
%\begin{displaymath}
$    p_R \defeq \prod_{u\in R}p_u\prod_{u\in\neigh{S}\setminus R}(1-p_u)$.
%\end{displaymath}

It is important to note that the process through which nodes arrive in the
second stage is \emph{not} an influence process.  The nodes in the second stage
arrive if they are willing to spread information in exchange for a unit of the
budget. Only when they have arrived does the influence process occur. This
process is encoded in the influence function and occurs after the influence
maximization stage without incentivizing nodes along the propagation path.  In
general, the idea of a two-stage (or in general, multi-stage) approach is to
use the nodes who arrive in the first stage to recruit influential users who
can be incentivized to spread information. In standard influence maximization,
the nodes who are not in the core set do not receive incentives to propagate
information, and cascades tend to die off
quickly~\cite{YC10,BHMW11,GWG12,CADKL14}.\newline

\noindent \textbf{Influence functions.}
In this paper we focus on \emph{linear} (or additive) influence models:
in these models the value of a subset of nodes can be expressed as
a weighted sum of their individual influence.  One important example of such
models is the \emph{voter model} \cite{richardson} used to represent the spread
of opinions in a social network: at each time step, a node adopts an opinion
with a probability equal to the fraction of its neighbors sharing this opinion
at the previous time step. Formally, this can be written as a discrete-time
Markov chain over opinion configurations of the network.  In this model
influence maximization amounts to ``converting'' the optimal subset of nodes to
a given opinion at the initial time so as to maximize the number of converts
after a given period of time.  Remarkably, a simple analysis shows that under
this model, the influence function $f$ is additive:
\begin{equation}\label{eq:voter}
    \forall S\subseteq V,\; f(S) = \sum_{u\in S} w_u
\end{equation}
where $w_u, u\in V$ are weights which can be easily computed from the powers of
the transition matrix of the Markov chain. This observation led to the
development of fast algorithms for influence maximization under the voter
model~\cite{even-dar}.\newline

\noindent \textbf{\textsc{NP}-Hardness.} In contrast to standard influence maximization, adaptive seeding is already \textsc{NP}-Hard even for the simplest influence functions such as $f(S) = |S|$ and when all probabilities are one.  We discuss this in Appendix~\ref{sec:alg-proofs}.
%In the case when $f(S)=|S|$ and all probabilities equal one, the decision problem is whether given a budget $k$ and target value $\ell$ there exists a subset of $X$ of size $k-t$ which yields a solution with expected value of $\ell$ using $t$ nodes in $\mathcal{N}(X)$.  It is easy to see that this problem is \textsc{NP}-hard by reduction from \textsc{Set-Cover}.
%This is equivalent to deciding whether there are $k-t$ nodes in $X$ that have $t$ neighbors in $\mathcal{N}(X)$.  To see this is \textsc{NP}-hard, consider reducing from \textsc{Set-Cover} where there is one node $i$ for each input set $T_i$, $1\leq i\leq n$, with $\neigh{i}= T_i$ and integers $k,\ell$, and the output is ``yes'' if there is a family of $k$ sets in the input which cover at least $\ell$ elements, and ``no'' otherwise.

\section{Non-adaptive Optimization}
\label{sec:adaptivity}
The challenging aspect of the adaptive seeding problem expressed in
Equation~\ref{eq:problem} is its adaptivity: a seed set must be selected during
the first stage such that in expectation a high influence value can be reached
when adaptively selecting nodes on the second stage.  A standard approach in 
stochastic optimization for overcoming this challenge is to use sampling to
estimate the expectation of the influence value reachable on the second stage.
However, as will be discussed in Section~\ref{sec:experiments}, this approach
quickly becomes infeasible even with modest size graphs.

In this section we develop an approach which avoids sampling and allows
designing adaptive seeding algorithms that can be applied to large graphs.  We
show that for additive influence functions one can optimize a relaxation of the
problem which we refer to as the \emph{non-adaptive} version of the problem.
After defining the non-adaptive version, we show in sections~\ref{sec:gap} that
the optimal solution for the non-adaptive version is an upper bound on the
optimal solution of the adaptive seeding problem.  We then argue in
Section~\ref{sec:round} that any solution to the non-adaptive version of the
problem can be converted to an adaptive solution, losing an arbitrarily small
factor in the approximation ratio.  Together, this implies that one can design
algorithms for the non-adaptive problem instead, as we do in
Section~\ref{sec:algorithms}.\newline 
      
%
%, namely \emph{non-adaptive}. As we will discuss in
%Sections~\ref{sec:gap} and~\ref{sec:round}, these non-adaptive policies can be
%used as tool to construct adaptive solutions.
%

\noindent \textbf{Non-adaptive policies.}  We say that a policy is \emph{non-adaptive} if it selects a set of nodes $S
\subseteq X$ to be seeded in the first stage and a vector of probabilities
$\mathbf{q}\in[0,1]^n$, such that each neighbor $u$ of $S$ which realizes is
included in the solution independently with probability $q_u$.  The constraint
will now be that the budget is only respected in expectation, \emph{i.e.}  $|S|
+ \textbf{p}^T\textbf{q} \leq k$. Formally the optimization problem for
non-adaptive policies can be written as:
\begin{equation}\label{eq:relaxed}
    \begin{split}
    %&\max_{}
        \max_{\substack{S\subseteq X\\\textbf{q}\in[0,1]^n} }& \;
    \sum_{R\subseteq\neigh{X}} \Big (\prod_{u\in R} p_uq_u\prod_{u\in\neigh{X}\setminus
    R}(1-p_uq_u) \Big )
 f(R)\\
    \text{s.t. } & \; |S|+\textbf{p}^T\textbf{q} \leq k,\;
q_u \leq \mathbf{1}\{u\in\neigh{S}\}
\end{split}
\end{equation}
%$\textbf{p}\odot \textbf{q}$ denotes componentwise multiplication and $\textbf{q}_R$ denotes the positive entries of $\textbf{q}$ on nodes in $R$: 
%
%\begin{displaymath}
%    \textbf{p}\circ \textbf{q}_R = \prod_{u\in R} p_uq_u\prod_{u\in\neigh{X}\setminus
%    R}(1-p_uq_u)
%\end{displaymath}
%
%\begin{displaymath}
%    \Pr[R \ |\ \textbf{p}\odot \textbf{q} ] = \prod_{u\in R} p_uq_u\prod_{u\in\neigh{X}\setminus
%    R}(1-p_uq_u)
%\end{displaymath}
where we denote by $\mathbf{1}\{E\}$ the indicator variable of the event $E$.
Note that because of the condition $q_u\leq \mathbf{1}\{u\in\neigh{S}\}$,
the summand associated with $R$ in \eqref{eq:relaxed} vanishes whenever $R$
contains $u\in\neigh{X}\setminus\neigh{S}$. Hence, the summation is restricted
to $R\subseteq\neigh{S}$ as in \eqref{eq:problem}.

%The relaxed non-adaptive optimization \eqref{eq:relaxed} problem can be written
%more concisely using the \emph{multi-linear extension} of $f$:
%\begin{displaymath}
%    \forall p\in[0,1]^m,\; F(p)
%    \defeq\mathbb{E}_{p_R}\big[f(R)\big]
%    =\sum_{R\subseteq\neigh{X}} p_R f(R)
%\end{displaymath}
%This \emph{extension by expectation} is known to present cross-convecity
%properties which can exploited in maximization problems when $f$ is
%a submodular function \cite{dughmi, vondrak}. Using this definition, \eqref{eq:relaxed}
%can be re-written as:
%\begin{equation}\label{eq:relaxed-multi}
%    \begin{split}
%    &\max_{S\subseteq X} \max_{q\in[0,1]^n}F(p\circ q)\\
%    &\text{s.t. }|S|+\sum_{u\in\neigh{X}}p_uq_u \leq k,\;
%q_u \leq \mathbf{1}_{\{u\in\neigh{S}\}}
%\end{split}
%\end{equation}
%
%When $f$ is an ìnfluence function from the triggering model, it has been proved
%in \cite{singer} that adaptive seeding has a bounded \emph{adaptivity gap}:
%denoting by $OPT$ the optimal solution of \eqref{eq:problem} and by $OPT_{NA}$
%the optimal solution of \eqref{eq:relaxed}, we have $OPT\leq\alpha OPT_{NA}$.
%This inequality justifies using non-adaptive policies to approximate solutions
%to the adaptive seeding problem.

\subsection{Adaptivity Gap}\label{sec:gap}

We will now justify the use of non-adaptive strategies by showing that the
optimal solution for this form of non-adaptive strategies yields a higher value
than adaptive ones.  For brevity, given a probability vector $\pi\in[0,1]^m$ we
write:
\begin{equation}\label{eq:multi}
        F(\pi) \defeq
        \sum_{R\subseteq\neigh{X}}\left(\prod_{u\in
        R}\pi_u\prod_{u\in\neigh{X}\setminus R}(1-\pi_u)\right)
    f(R)
\end{equation}
as well as $\textbf{p}\otimes \textbf{q}$ to denote the component-wise
multiplication between vectors $\textbf{p}$ and $\textbf{q}$.  Finally, we write
$\mathcal{F}_{A} \defeq \{S \subseteq X : |S|\leq k\}$, and $\mathcal{F}_{NA}
\defeq\{(S,\textbf{q}), |S|+\textbf{p}^T\textbf{q} \leq k, q_u \leq
\mathbf{1}_{\{u\in\neigh{S}\}}\}$ to denote the feasible regions of the
adaptive and non-adaptive problems, respectively.

%this form of non-adap
%We already know that the adaptivity gap is bounded for a general class of
%influence function. For adaptive functions, we get a stronger result:
%\emph{relaxed-non adaptive policies are stronger than non-adaptive policies}.
%
%
\begin{proposition}\label{prop:gap}
For additive functions given by \eqref{eq:voter}, the value of the optimal
adaptive policy is upper bounded by the optimal non-adaptive policy:
    \begin{displaymath}
        \begin{aligned}[t]
            &\max_{S\subseteq X} \sum_{R\subseteq\neigh{S}} p_R
            \max_{\substack{T\subseteq R\\|T|\leq k-|S|}}f(T)\\
            &\text{s.t. }S \in \mathcal{F}_{A}
        \end{aligned}
        \leq
        \begin{aligned}[t]
            &\max_{\substack{S\subseteq X\\\textbf{q}\in[0,1]^n}}
            F(\mathbf{p}\otimes\mathbf{q})\\
            &\text{s.t. } (S,\textbf{q}) \in \mathcal{F}_{NA}
                    \end{aligned}
    \end{displaymath}
\end{proposition}

%\begin{proposition}\label{prop:gap}
%Let $\mathcal{F}_{A} := \{T \subseteq X : |T|\leq k\}$, $\mathcal{F}_{NA} :=\{(T,\textbf{q}), |S|+\textbf{p}^T\textbf{q} \leq k, q_u \leq \mathbf{1}_{\{u\in\neigh{S}\}}\}$.
%For additive functions of the form given by \eqref{eq:voter}, non-adaptive
%    policies are stronger than adaptive policies:
%    \begin{displaymath}
%        \begin{aligned}[t]
%            &\max_{S\subseteq X} \sum_{R\subseteq\neigh{S}} p_R
%            \max_{\substack{T\subseteq R\\|T|\leq k-|S|}}f(T)\\
%            &\text{s.t. }|S|\leq k
%        \end{aligned}
%        \leq
%        \begin{aligned}[t]
%            &\max_{S\subseteq X} \max_{q\in[0,1]^n}F(p\circ q)\\
%            &\text{s.t. }|S|+\sum_{u\in\neigh{X}}p_uq_u \leq k,\;
%            q_u \leq \mathbf{1}_{\{u\in\neigh{S}\}}
%        \end{aligned}
%    \end{displaymath}
%\end{proposition}

The proof of this proposition can be found in Appendix~\ref{sec:ad-proofs} and
relies on the following fact: the optimal adaptive policy can be written as
a feasible non-adaptive policy, hence it provides a lower bound on the value of
the optimal non-adaptive policy.

\subsection{From Non-Adaptive to Adaptive Solutions}\label{sec:round}

From the above proposition we now know that optimal non-adaptive solutions have
higher values than adaptive solutions. Given a non-adaptive solution
$(S,\mathbf{q})$, a possible scheme would be to use $S$ as an adaptive
solution.  But since $(S, \mathbf{q})$ is a solution to the non-adaptive
problem, Proposition~\ref{prop:gap} does not provide any guarantee on how well
$S$ performs as an adaptive solution.

However, we show that from a non-adaptive solution $(S,\mathbf{q})$, we can
obtain a lower bound on the adaptive value of $S$, that is, the expected
influence attainable in expectation over all possible arrivals of neighbors of
$S$. Starting from $S$, in every realization of neighbors $R$, sample every
node $u \in R \cap \mathcal{N}(S)$ with probability $q_{u}$, to obtain a random
set of nodes $I_R \subseteq R \cap S$. $(S, \mathbf{q})$ being a non-adaptive
solution, it could be that selecting $I_R$ exceeds our budget. Indeed, the only
guarantee that we have is that $|S| + \mathbb{E}\big[|I_R|\big]\leq k$. As
a consequence, an adaptive solution starting from $S$ might not be able to
select $I_R$ on the second stage.

%execute an adaptive policy by selecting a set $S$ in the first stage, and then in the second stage, while there is still budget remaining, select every neighbor $u$ of $S$ that realized with probability $q_{u}$

%To do so, one needs to prove that in expectation over all the realizations, the average values obtainable in the second stage, are close to the value of the non-adaptive objective evaluated over $(S,\mathbf{q})$.
 
%the objective function of the adaptive problem \eqref{eq:relaxed}. However, the set resulting from the sampling might exceed the budget on the second stage, hence preventing us from directly obtaining a feasible adaptive solution.    
%Given a a non-adaptive policy $(S,\mathbf{q})$ one can execute an adaptive policy by selecting a set $S$ in the first stage, and then in the second stage, while there is still budget remaining, select every neighbor $u$ of $S$ that realized with probability $q_{u}$.  
%In order to use the above lemma, one needs to show that in almost all realizations   
%non-adaptive policy

%Fortunately, by using contention resolution schemes~\cite{vondrak} given a feasible solution $\mathbf{q}$ to the non-adaptive version, one can compute a \emph{feasible} random set $I$, such that:
%\begin{equation}\label{eq:cr}
%\mathbb{E}\big[f(I)\big] 
%\geq (1-\varepsilon) F(\mathbf{q}) 
%\end{equation}

Fortunately, the probability of exceeding the budget is small enough and with
high probability $I_R$ will be feasible. This is exploited in \cite{vondrak} to
design a randomized rounding method with approximation guarantees. These
rounding methods are called \emph{contention resolution schemes}.
%More precisely, we note that once the set $S$ is fixed, the feasibility constraint of problem~\eqref{eq:relaxed} becomes a single Knapsack constraint, for which \cite{vondrak} constructs a $(1-\varepsilon, 1-\varepsilon)$-balanced contention resolution scheme.
Theorem~1.3 of this paper gives us a contention resolution scheme which will
compute from $\mathbf{q}$ and for any realization $R$ a \emph{feasible} set
$\tilde{I}_R$, such that:
\begin{equation}
    \label{eq:cr}
    \mathbb{E}_R\big[f(\tilde{I}_R)\big] \geq (1-\varepsilon) F(\mathbf{q})
\end{equation}
What this means is that starting from a non-adaptive solution $(S,\mathbf{q})$,
there is a way to construct a random \emph{feasible} subset on the second stage
such that in expectation, this set attains almost the same influence value as
the non-adaptive solution. Since the adaptive solution starting from $S$ will
select optimally from the realizations $R\subseteq\neigh{S}$,
$\mathbb{E}_R[f(\tilde{I}_R)]$ provides a lower bound on the adaptive value of
$S$ that we denote by $A(S)$.

More precisely, denoting by $\text{OPT}_A$ the optimal value of the adaptive
problem~\eqref{eq:problem}, we have the following proposition whose proof can
be found in Appendix~\ref{sec:ad-proofs}.
\begin{proposition}\label{prop:cr}
    Let $(S,\textbf{q})$ be an $\alpha$-approximate solution to the
    non-adaptive problem \eqref{eq:relaxed}, then $\mathrm{A}(S) \geq \alpha
    \mathrm{OPT}_A$.
\end{proposition}

\section{Algorithms}
\label{sec:algorithms}
Section~\ref{sec:adaptivity} shows that the adaptive seeding problem reduces to
the non-adaptive problem.  We will now discuss two approaches to construct
approximate non-adaptive solutions. The first is an LP-based approach, and the
second is a combinatorial algorithm. Both approaches have the same $(1-1/e)$
approximation ratio, which is then translated to a $(1-1/e)$ approximation
ratio for the adaptive seeding problem~\eqref{eq:problem} via
Proposition~\ref{prop:cr}. As we will show in Section~\ref{sec:experiments},
both algorithms have their advantages and disadvantages in terms of
scalability.

\subsection{An LP-Based Approach}
\label{sec:lp}
Note that due to linearity of expectation, for a linear function $f$ of the
form given by \eqref{eq:voter} we have:
\begin{equation}\label{eq:multi-voter}
    \begin{split}
        F(\textbf{p}) 
        &=\mathbb{E}_{R}\big[f(R)\big]
        =\mathbb{E}_{R}\Bigg[\sum_{u\in\neigh{X}}w_u\mathbf{1}_{\{u\in
        R\}}\Bigg]\\
        &=\sum_{u\in\neigh{X}}w_u\mathbb{P}[u\in R]
        =\sum_{u\in\neigh{X}}p_uw_u
    \end{split}
\end{equation}

Thus, the non-adaptive optimization problem \eqref{eq:relaxed} can be written as:
\begin{displaymath}
    \begin{split}
        \max_{\substack{S\subseteq X\\\mathbf{q}\in[0,1]^n} } 
        & \sum_{u\in\neigh{X}}p_uq_uw_u\\
        \text{s.t. } & |S|+ \textbf{p}^T\textbf{q} \leq k,\;
                     q_u \leq \mathbf{1}\{u\in\neigh{S}\}
    \end{split}
\end{displaymath}

The choice of the set $S$ can be relaxed by introducing a variable
$\lambda_v\in[0,1]$ for each $v\in X$. We obtain the following
LP for the adaptive seeding problem:
\begin{equation}\label{eq:lp}
    \begin{split}
        \max_{\substack{\mathbf{q}\in[0,1]^n\\\boldsymbol\lambda\in[0,1]^m}}
        & \;\sum_{u\in\neigh{X}}p_uq_uw_u\\
        \text{s.t. } & \sum_{v\in X}\lambda_v+\textbf{p}^T\textbf{q} \leq k,\;
        q_u \leq \sum_{v\in\neigh{u}} \lambda_v
\end{split}
\end{equation}

An optimal solution to the above problem can be found in polynomial time using
standard LP-solvers.  The solution returned by the LP is \emph{fractional}, and
requires a rounding procedure to return a feasible solution to our problem,
where $S$ is integral.  To round the solution we use the pipage rounding
method~\cite{pipage}.  We defer the details to Appendix~\ref{sec:lp-proofs}.

\begin{lemma}
    For \mbox{\textsc{AdaptiveSeeding-LP}} defined in \eqref{eq:lp}, any fractional solution $(\boldsymbol\lambda, \mathbf{q})\in[0,1]^m\times[0,1]^n$ can be rounded to an integral solution $\bar{\boldsymbol\lambda} \in \{0,1\}^{m}$ s.t. $(1-1/e) F(\mathbf{p}\circ\mathbf{q}) \leq A(\bar{\lambda})$ in $O(m + n)$ steps.
\end{lemma}

\subsection{A Combinatorial Algorithm}
\label{sec:comb}

In this section, we introduce a combinatorial algorithm with an identical
approximation guarantee to the LP-based approach. However, its running time,
stated in Proposition~\ref{prop:running_time} can be better than the one given
by LP solvers depending on the relative sizes of the budget and the number of
nodes in the graph. Furthermore, as we discuss at the end of this section, this
algorithm is amenable to parallelization. 

The main idea is to reduce the problem to a monotone submodular maximization
problem and apply a variant of the celebrated greedy
algorithm~\cite{nemhauser}.  
%This property is quite a remarkable feature of
%linear influence models; for influence models such as independent cascade and
%linear threshold, the adaptive seeding problem does not reduce to submodular
%maximization, and a completely different approach is required~\cite{singer}. 
In contrast to standard influence maximization, the submodularity of the
non-adaptive seeding problem is not simply a consequence of properties of the
influence function; it also strongly relies on the combinatorial structure of
the two-stage optimization. 

Intuitively, we can think of our problem as trying to find a set $S$ in the
first stage, for which the nodes that can be seeded on the second stage have
the largest possible value.  To formalize this, for a budget $b\in\reals^+$
used in the second stage and a set of neighbors $T\subseteq\mathcal{N}(X)$, we
will use $\mathcal{O}(T,b)$ to denote the solution to:
\begin{equation}\label{eq:knap}
    \begin{split}
        \mathcal{O}(T,b)\defeq
        \max_{\textbf{q}\in[0,1]^n} & \sum_{u\in\neigh{X} \cap T} p_uq_uw_u\\
                            \text{s.t. } & \mathbf{p}^T\mathbf{q}\leq b
                            %\text{ and } q_u=0\text{ if}u\notin T
\end{split}
\end{equation}

The optimization problem \eqref{eq:relaxed} for
non-adaptive policies can now be written as:
\begin{equation}\label{eq:sub}
        \max_{S\subseteq X} \; \mathcal{O}\big(\neigh{S},k-|S|\big)
        \quad \text{s.t. } |S|\leq k
\end{equation}

We start by proving in Proposition~\ref{prop:sub} that for fixed $t$,
$\mathcal{O}(\neigh{\cdot}, t)$ is submodular. This proposition relies on
lemmas~\ref{lemma:nd} and~\ref{lemma:sub} about the properties of
$\mathcal{O}(T,b)$.

\begin{lemma}\label{lemma:nd}
    Let $T \subseteq \mathcal{N}(X)$ and $x \in \mathcal{N}(X)$, then
    $\mathcal{O}(T\cup\{x\},b)-\mathcal{O}(T,b)$ is
    non-decreasing in $b$.
\end{lemma}

The proof of this lemma can be found in Appendix~\ref{sec:comb-proofs}. The main
idea consists in writing:
\begin{multline*}
    \mathcal{O}(T\cup\{x\},c)-\mathcal{O}(T\cup\{x\},b)=\int_b^c\partial_+\mathcal{O}_{T\cup\{x\}}(t)dt
\end{multline*}
where $\partial_+\mathcal{O}_T$ denotes the right derivative of
$\mathcal{O}(T,\cdot)$. For a fixed $T$ and $b$, $\mathcal{O}(T,b)$ defines
a fractional Knapsack problem over the set $T$. Knowing the form of the optimal
fractional solution, we can verify that
$\partial_+\mathcal{O}_{T\cup\{x\}}\geq\partial_+\mathcal{O}_T$ and obtain:
\begin{multline*}
    \mathcal{O}(T\cup\{x\},c)-\mathcal{O}(T\cup\{x\},b)\geq 
    \mathcal{O}(T,c)-\mathcal{O}(T,b)
\end{multline*}

\begin{lemma}\label{lemma:sub}
    For any $b\in\reals^+$, $\mathcal{O}(T,b)$ is submodular in $T$, $T\subseteq\neigh{X}$.
\end{lemma}

The proof of this lemma is more technical. For $T\subseteq\neigh{X}$ and $x,
y\in\neigh{X}\setminus T$, we need to show that:
    \begin{displaymath}
        \mathcal{O}(T\cup\{x\},b)-\mathcal{O}(T,b)\geq
        \mathcal{O}(T\cup\{y, x\},b)-\mathcal{O}(T\cup\{y\},b)
    \end{displaymath}
    This can be done by partitioning the set $T$ into ``high value
    items'' (those with weight greater than $w_x$) and ``low value items'' and
    carefully applying Lemma~\ref{lemma:nd} to the associated subproblems.
    The proof is in Appendix~\ref{sec:comb-proofs}.

Finally, Lemma~\ref{lemma:sub} can be used to show Proposition~\ref{prop:sub}
whose proof can be found in Appendix~\ref{sec:comb-proofs}.

\begin{proposition}\label{prop:sub}
    Let $b\in\mathbf{R}^+$, then $\mathcal{O}(\neigh{S},b)$ is monotone and
    submodular in $S$, $S\subseteq X$.
\end{proposition}

We can now use Proposition~\ref{prop:sub} to reduce \eqref{eq:sub} to a monotone submodular maximization problem. First, we note that~\eqref{eq:sub} can be rewritten:
\begin{equation}\label{eq:sub-mod}
    \max_{\substack{S\subseteq X\\ t \in \mathbb{N}}} \; \mathcal{O}\big(\neigh{S},t\big)
        \quad\text{s.t. } |S| + t\leq k
\end{equation}

Intuitively, we fix $t$ arbitrarily so that the maximization above becomes a submodular maximization problem with fixed budget $t$. We then optimize over the value of $t$. Combining this observation with the greedy algorithm for monotone submodular maximization~\cite{nemhauser}, we obtain Algorithm~\ref{alg:comb}, whose performance guarantee is summarized in Proposition~\ref{prop:main_result}.

\begin{algorithm}
    \caption{Combinatorial algorithm}
    \label{alg:comb}
    \algsetup{indent=2em}
    \begin{algorithmic}[1]
        \STATE $S\leftarrow \emptyset$
        \FOR{$t=1$ \TO $k-1$}
            \STATE $S_t\leftarrow \emptyset$
            \FOR{$i=1$ \TO $k-t$}
                \STATE $x^*\leftarrow\argmax_{x\in
                X\setminus S_t}\mathcal{O}(\neigh{S_t\cup\{x\}},t)
                -\mathcal{O}(\neigh{S_t},t)$\label{line:argmax}
                \STATE $S_t\leftarrow S_t\cup\{x^*\}$
            \ENDFOR
            \IF{$\mathcal{O}(\neigh{S_t},t)>\mathcal{O}(\neigh{S},k-|S|)$}
                \STATE $S\leftarrow S_t$
            \ENDIF
        \ENDFOR
        \RETURN $S$
    \end{algorithmic}
\end{algorithm}

\begin{proposition}\label{prop:main_result}
    Let $S$ be the set computed by Algorithm~\ref{alg:comb} and let us denote
    by $\mathrm{A}(S)$ the value of the adaptive policy selecting $S$ on the first
    stage. Then $\mathrm{A}(S) \geq (1-1/e)\mathrm{OPT}_A$.
\end{proposition}

%\subsubsection{Scalability}

\noindent\textbf{Parallelization.}  The algorithm described above considers all
possible ways to split the seeding budget between the first and the second
stage.  For each possible split $\{(t,k-t)\}_{t=1\ldots,k-1}$, the algorithm
computes an approximation to the optimal non adaptive solution that uses $k-t$
nodes in the first stage and $t$ nodes in the second stage, and returns the
solution for the split with the highest value (breaking ties arbitrarily).
This process can be trivially parallelized across $k-1$ machines, each
performing a computation of a single split.  With slightly more effort, for any
$\epsilon>0$ one can parallelize over $\log_{1+\epsilon}n$ machines at the cost
of losing a factor of $\epsilon$ in the approximation guarantee (see Appendix~\ref{sec:para} for details).\newline

\noindent \textbf{Implementation in MapReduce.} While the previous paragraph
describes how to parallelize the outer \texttt{for} loop of
Algorithm~\ref{alg:comb}, we note that its inner loop can also be parallelized
in the MapReduce framework. Indeed, it corresponds to the greedy algorithm
applied to the function $\mathcal{O}\left(\neigh{\cdot}, t\right)$. The
\textsc{Sample\&Prune} approach successfully applied in \cite{mr} to obtain
MapReduce algorithms for various submodular maximizations can also be applied
to Algorithm~\ref{alg:comb} to cast it in the MapReduce framework. The details
of the algorithm can be found in Appendix~\ref{sec:mr}.
\newline

%A slightly more sophisticated approach is to consider only $\log n$ splits: $(1,k-1),(2,k-2),\ldots,(2^{\lfloor \log n \rfloor},1)$ and then select the best solution from this set.  It is not hard to see that in comparison to the previous approach, this would reduce the approximation guarantee by a factor of at most 2: if the optimal solution is obtained by spending $t$ on the first stage and $k-t$ in the second stage, then since $t \leq 2\cdot2^{\lfloor \log t \rfloor}$ the solution computed for $(2^{\lfloor \log t \rfloor}, k - 2^{\lfloor \log t \rfloor})$ will have at least half that value.  
%More generally, for any $\epsilon>0$ one can parallelize over $\log_{1+\epsilon}n$ machines at the cost of losing a factor of $(1+\epsilon)$ in the approximation guarantee.

\noindent \textbf{Algorithmic speedups.}  To implement Algorithm~\ref{alg:comb} efficiently, the computation of the $\argmax$ on line 5 must be dealt with carefully. $\mathcal{O}(\neigh{S_t\cup\{x\}},t)$ is the optimal solution to the fractional Knapsack problem~\eqref{eq:knap} with budget $t$ and can be computed in time $\min(\frac{t}{p_\text{min}},n)$ by iterating over the list of nodes in $\neigh{S_t\cup\{x\}}$ in decreasing order of the degrees. This decreasing order of $\neigh{S_t}$ can be maintained throughout the greedy construction of $S_t$ by:
\begin{itemize}
    \item ordering the list of neighbors of nodes in $X$ by decreasing order of the degrees when initially constructing the graph. This is responsible for a $O(n\log n)$ pre-processing time.
\item when adding node $x$ to $S_t$, observe that $\neigh{S_t\cup\{x\}} = \neigh{S_t}\cup\neigh{\{x\}}$. Hence, if $\neigh{S_t}$ and $\neigh{\{x\}}$ are sorted lists, then $\mathcal{O}(\neigh{S_t\cup\{x\}},t)$ can be computed in a single iteration of length $\min(\frac{t}{p_\text{min}},n)$ where the two sorted lists are merged on the fly.
\end{itemize}
As a consequence, the running time of line 5 is bounded from above by $m\min(\frac{t}{p_\text{min}},n)$. The two nested \textsf{for} loops are responsible for the additional $k^2$ factor. The running time of Algorithm~\ref{alg:comb} is summarized in Proposition~\ref{prop:running_time}.

\begin{proposition}\label{prop:running_time}
    Let $p_\text{min} =\min\{p_u,u\in\neigh{X}\}$, then Algorithm~\ref{alg:comb} runs in time 
    ${O\big(n\log n + k^2 m \min(\frac{k}{p_\text{min}},n)\big)}$.
\end{proposition}

\section{Experiments}
\label{sec:experiments}
In this section we validate the adaptive seeding approach through experimentation.
%In the sections above we showed how to design algorithms that have provable performance guarantees for adaptive seeding, which and even parallelizable.    
Specifically, we show that our algorithms for adaptive seeding obtain
significant improvement over standard influence maximization, that these
improvements are robust to changes in environment variables, and that our
approach is efficient in terms of running-time and scalable to large social
networks.

\subsection{Experimental setup}
We tested our algorithms on three types of datasets. Each of them allows us to
experiment on a different aspect of the adaptive seeding problem. The Facebook
Pages dataset that we collected ourselves has a central place in our
experiments since it is the one which is closet to actual applications of
adaptive seeding.

\textbf{Synthetic networks.} Using standard models of social networks we
generated large-scale graphs to model the social network.  To emulate the
process of users following a topic (the core set $X$) we sampled subsets of
nodes at random, and applied our algorithms on the sample and their neighbors.
The main advantage of these data sets is that they allow us to generate graphs
of arbitrary sizes and experiment with various parameters that govern the
structure of the graph.  The disadvantages are that users who follow a topic
are not necessarily random samples, and that social networks often have
structural properties that are not captured in generative models.

\textbf{Real networks.}  We used publicly available data sets of real social
networks available at \cite{snapnets}. As for synthetic networks, we used
a random sample of nodes to emulate users who follow a topic, which is the main
disadvantage of this approach.  The advantage however, is that such datasets
contain an entire network which allows testing different propagation
parameters.   
  
\textbf{Facebook Pages.}  We collected data from several Facebook Pages, each
associated with a commercial entity that uses the Facebook page to communicate
with its followers. For each page, we selected a post and then collected data
about the users who expressed interest (``liked'') the post and their friends.
The advantage of this data set is that it is highly representative of the
scenario we study here. Campaigns run on a social network will primarily target
users who have already expressed interests in the topic being promoted. The
main disadvantage of this method is that such data is extremely difficult to
collect due to the crawling restrictions that Facebook applies and gives us
only the 2-hop neighborhood around a post. This makes it difficult to
experiment with different propagation parameters. Fortunately, as we soon
discuss, we were able to circumvent some of the crawling restrictions and
collect large networks, and the properties of the voter influence model are
such that these datasets suffice to accurately account for influence
propagation in the graph.\newline

\begin{figure}[t]
    \centering
    \includegraphics[width=0.4\textwidth]{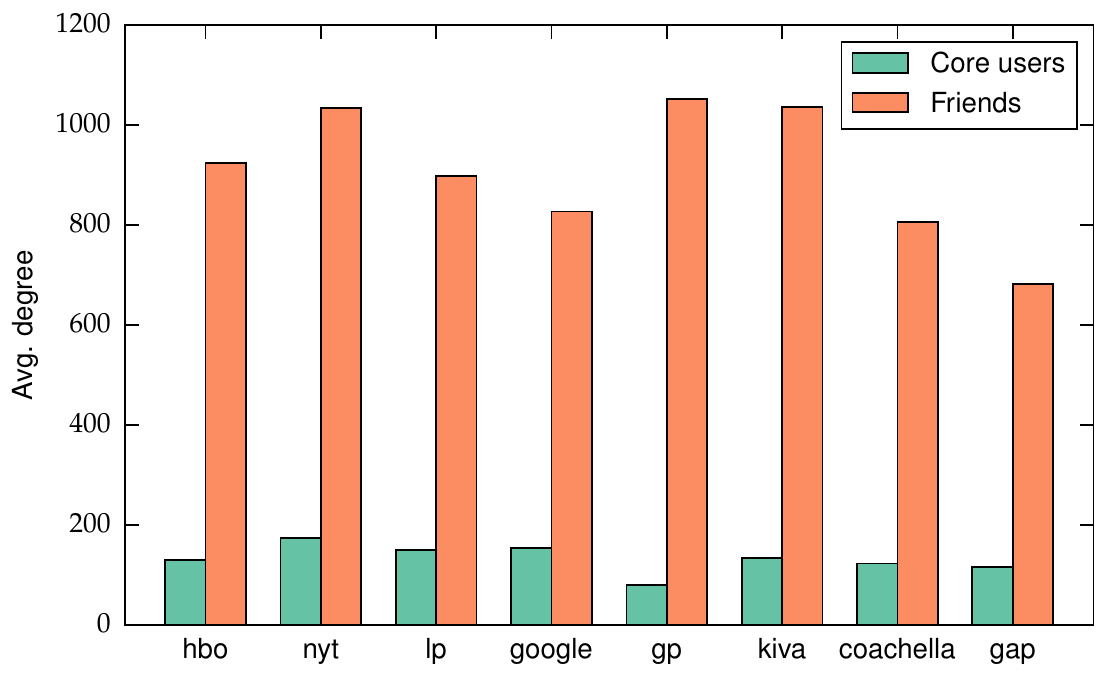}
    \caption{Comparison of the average degree of the core set users and the
    average degree of their friends.}
    \label{fig:paradox}
    \vspace{-10pt}
\end{figure}

\noindent\textbf{Data collection.}
We selected Facebook Pages in different verticals (topics). Each page is
operated by an institution or an entity whose associated Facebook Page is
regularly used for promotional posts related to this topic. On each of these
pages, we selected a recent post (posted no later than January 2014) with
approximately 1,000 \emph{likes}. The set of users who liked those posts
constitute our core set.  We then crawled the social network of
these sets: for each user, we collected her list of friends, and the degrees
(number of friends) of these friends.\newline

\noindent\textbf{Data description.} Among the several verticals we collected,
we select eight of them for which we will report our results. We obtained
similar results for the other ones.  Table~\ref{tab:data} summarizes statistics
about the selected verticals.  We note that depending on the privacy settings
of the core set users, it was not always possible to access their list of
friends.  We decided to remove these users since their ability to spread
information could not be readily determined. This effect, combined with various
errors encountered during the data collection, accounts for an approximate 15\%
reduction between the users who liked a post and the number of users in the
datasets we used.  Following our discussion in the introduction, we observe
that on average, the degrees of core set users is much lower than the degrees of
their friends. This is highlighted on Figure~\ref{fig:paradox} and justifies
our approach.

\begin{table}[t]
    \small
    \centering
    \setlength{\tabcolsep}{3pt}
    \begin{tabular}{llrr}
    \toprule
    Vertical & Page & $m$ & $n$ \\%& $S$ & $F$\\
    \midrule
    Charity & Kiva & 978 & 131334 \\%& 134.29 & 1036.26\\
    Travel & Lonely Planet & 753 & 113250 \\%& 150.40 & 898.50\\
    %Public Action & LaManifPourTous & 1041 & 97959 \\%& 94.10 & 722.02\\
    Fashion & GAP & 996 & 115524 \\%& 115.99 & 681.98\\
    Events & Coachella & 826 & 102291 \\%& 123.84 & 870.16\\
    Politics & Green Party & 1044 & 83490 \\%& 79.97 & 1053.25\\
    Technology & Google Nexus & 895 & 137995 \\%& 154.19 & 827.84\\
    News & The New York Times & 894 & 156222 \\%& 174.74 & 1033.94 \\
    %Consumption & Peet's & 776 & 56268 \\%& 72.51 & 520.47\\
    Entertainment & HBO & 828 & 108699 \\%& 131.28 & 924.09\\
    \bottomrule
\end{tabular}
\caption{Dataset statistics. $m$: number of users in the core set, $n$: number
of friends of core set users.}
%$S$: avg. degree of an initial user, $F$: avg. degree of a friend of an initial user.}
\label{tab:data}
    \vspace{-10pt}
\end{table}

\subsection{Performance of Adaptive Seeding}
\label{sec:performance} For a given problem instance with a budget of $k$ we
applied the adaptive seeding algorithm (the combinatorial version). Recall from
Section~\ref{sec:model} that performance is defined as the expected influence
that the seeder can obtain by optimally selecting users on the second stage,
where \emph{influence} is defined as the sum of the degrees of the selected
users. We tested our algorithm against the following benchmarks:

\begin{itemize}
    \item \emph{Random Node} (\textsf{RN}): we randomly select $k$ users from
        the core set.  This is a typical benchmark in comparing influence
        maximization algorithms~\cite{KKT03}.
   \item \emph{Influence Maximization} (\textsf{IM}): we apply the optimal
       influence maximization algorithm on the core set.  This is the naive
       application of influence maximization. For the voter model, when the
       propagation time is polynomially large in the network size, the optimal
       solution is to simply take the $k$ highest degree nodes~\cite{even-dar}.
       We study the case of bounded time horizons in Section~\ref{sec:inf}.
    \item \emph{Random Friend} (\textsf{RF}): we implement a naive two-stage approach:
        randomly select $k/2$ nodes from the core set, and for each
        node select a random neighbor (hence spending the budget of $k$ rewards
        overall).  This method was recently shown to outperform standard
        influence maximization when the core set is random~\cite{LS13}.
\end{itemize}

\begin{figure*}[t]
    \centerline{\includegraphics[width=0.99\textwidth]{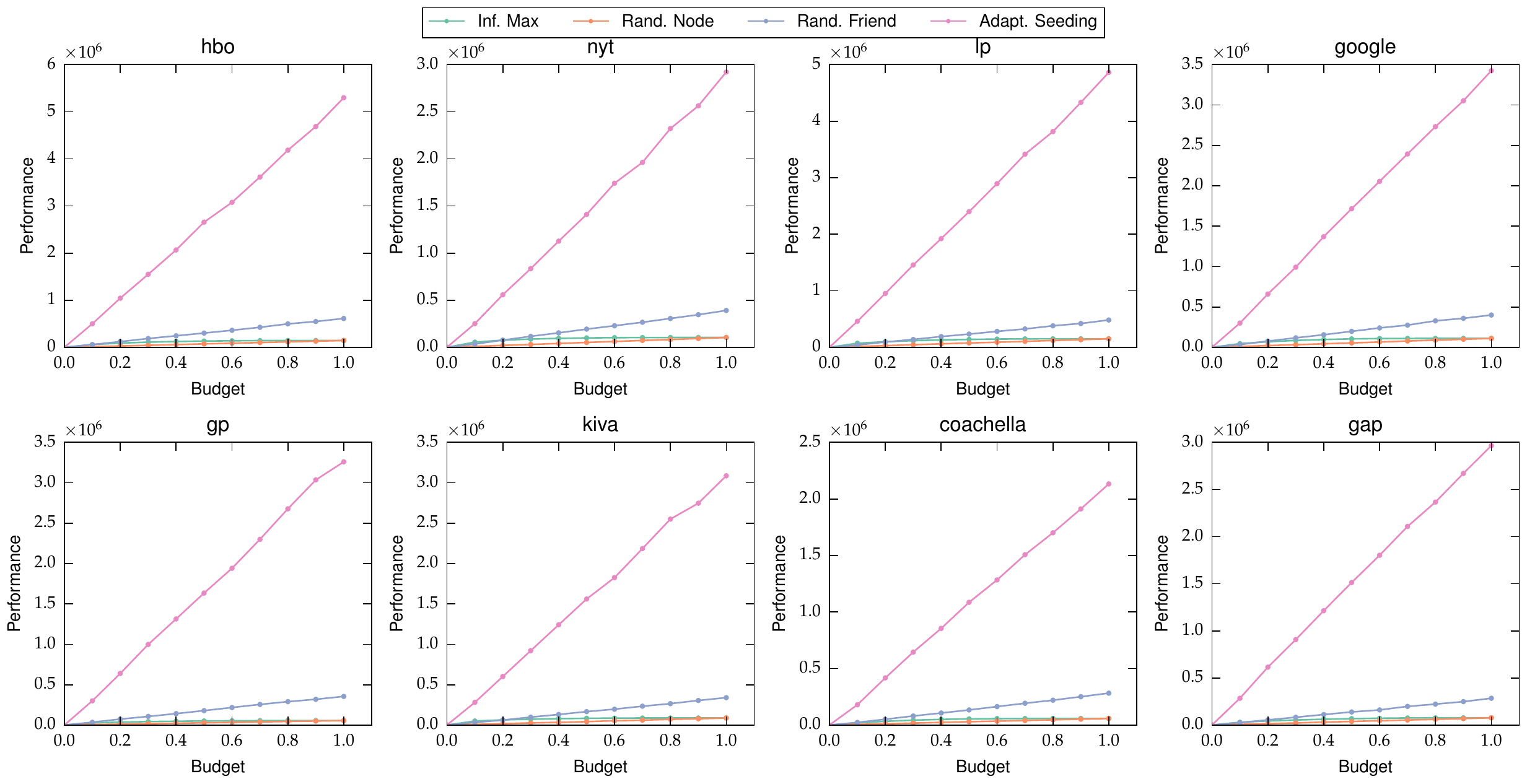}}
    \vspace{-5pt}
    \caption{\small{Performance of adaptive seeding compared to other influence
    maximization approaches. The horizontal axis represents the budget used as
a fraction of the size of the core set. The vertical axis is the
expected influence reachable by optimally selecting nodes on the second stage.}}
    \label{fig:performance}
    \vspace{-10pt}
\end{figure*}

\subsection{Performance on Facebook Pages} Figure~\ref{fig:performance}
compares the performance of \emph{adaptive seeding}, our own approach, to the
afore-mentioned approaches for all the verticals we collected.  In this first
experiment we made simplifying assumptions about the parameters of the model.
The first assumption is that all probabilities in the adaptive seeding model
are equal to one. This implicitly assumes that every friend of a user who
followed a certain topic is interested in promoting the topic given a reward.
Although this is a strong assumption that we will revisit, we note that the
probabilities can be controlled to some extent by the social networking service
on which the campaign is being run by showing prominently the campaign material
(sponsored links, fund-raising banners, etc.).  The second assumption is that
the measure of influence is the sum of the degrees of the selected set.  This
measure is an appealing proxy as it is known that in the voter model, after
polynomially many time steps, the influence of each node is proportional to its
degree with high probability~\cite{even-dar}. Since the influence process
cannot be controlled by the designer, the assumption is often that the
influence process runs until it stabilizes (in linear thresholds and
independent cascades for example, the process terminates after a linear number
of steps~\cite{KKT03}).  We perform a set of experiments for different time
horizons in Section~\ref{sec:inf}.

\begin{figure}[t]
    \vspace{-10pt}
    \centerline{ \includegraphics[width=0.4\textwidth]{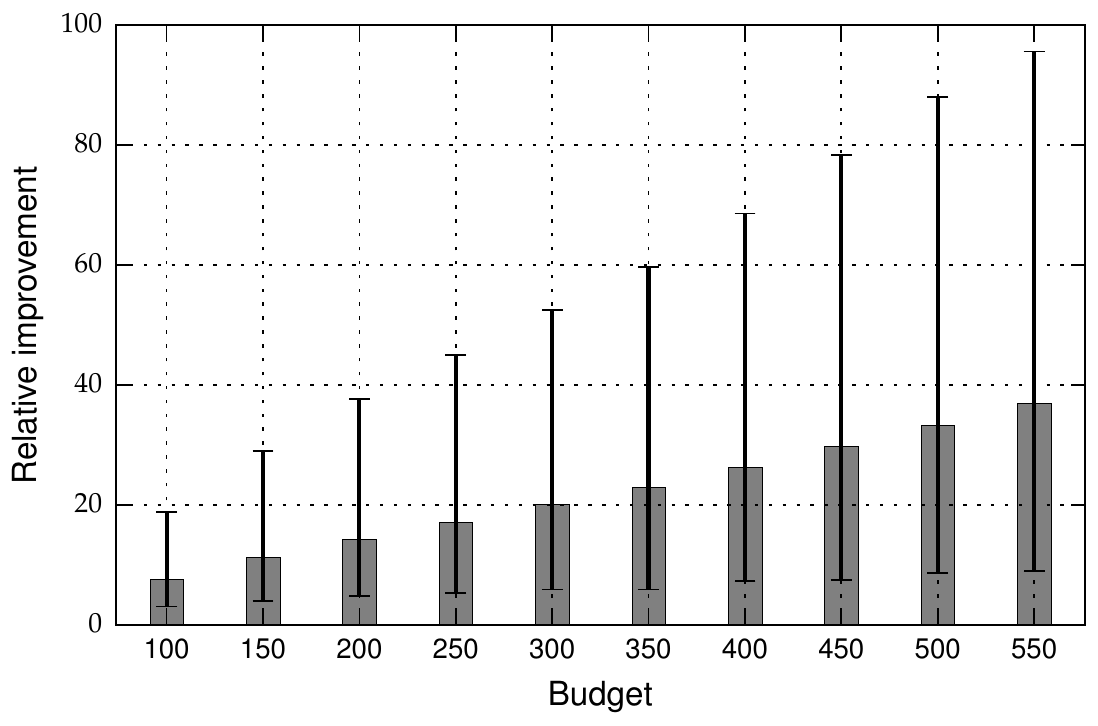} }
    \vspace{-10pt}
    \caption{Ratio of the performance of adaptive seeding to \textsf{IM}. Bars represents the mean improvement across all verticals, and the ``error bar'' represents the range of improvement across verticals.}
    \label{fig:compare}
    \vspace{-15pt}
\end{figure}

It is striking to see how well adaptive seeding does in comparison to other
methods.  Even when using a small budget (0.1 fraction of the core set, which
in these cases is about 100 nodes), adaptive seeding improves influence by
a factor of at least 10, across all verticals.  To confirm this, we plot the
relative improvements of \emph{adaptive seeding} over \textsf{IM} in aggregate
over the different pages.  The results are shown in Figure~\ref{fig:compare}.
This dramatic improvement is largely due to the friendship paradox phenomenon
that adaptive seeding leverages.  Returning to Figure~\ref{fig:performance}, it
is also interesting to note that the \textsf{RF} heuristic significantly
outperforms the standard \textsf{IM} benchmark.  Using the same budget, the
degree gain induced by moving from the core set to its neighborhood is such
that selecting at random among the core set users' friends already does better
than the best heuristic restricted only on the core set. Using \emph{adaptive
seeding} to optimize the choice of core set users based on their friends'
degrees then results in an order of magnitude increase over \textsf{RF},
consistently for all the pages.\newline

\subsection{The Effect of the Probabilistic Model}
\label{sec:robustness}

The results presented in Section~\ref{sec:performance} were computed assuming
the probabilities in the adaptive seeding model are one.  We now describe
several experiments we performed with the Facebook Pages data set that test the
advantages of adaptive seeding under different probability models.\newline 

%Estimating this probability is a research problem on its own; however,
%we note that it can be controlled to some extent by the social networking
%service on which the campaign is being run. By showing prominently the campaign
%material (sponsored links, fund-raising banners, etc.), the conversion rate can
%be increased beyond what would happen via regular word-of-mouth propagation.
%\newline

\begin{figure}[t]
    \begin{subfigure}[t]{0.23\textwidth}
    \includegraphics[scale=0.48]{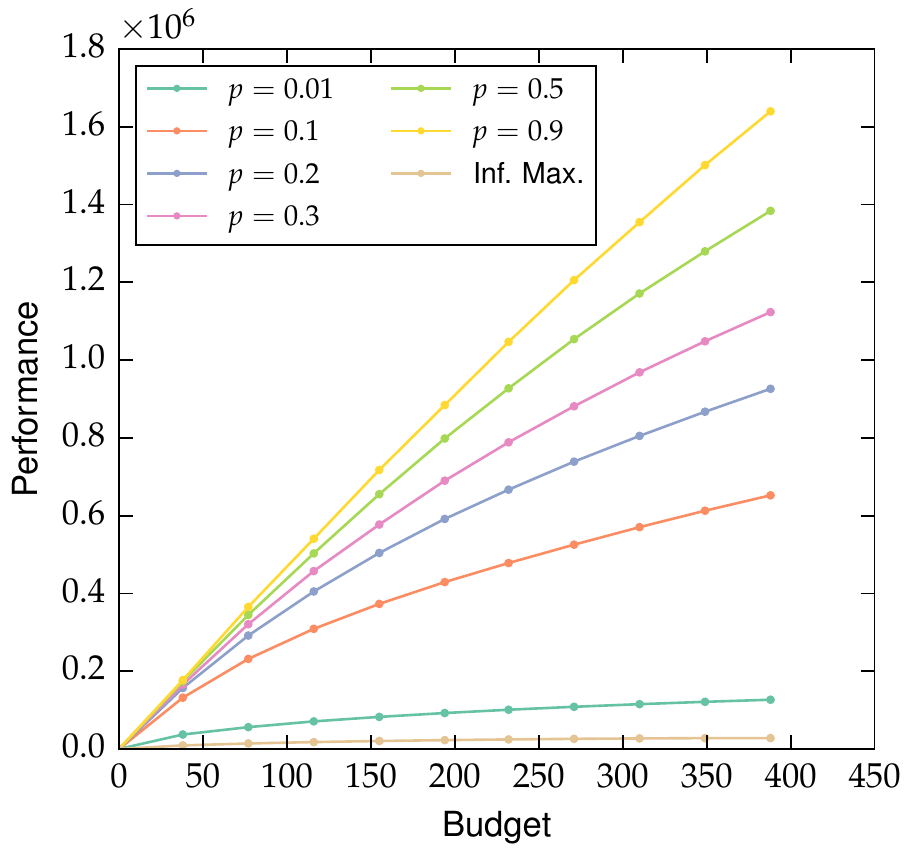}
    \vspace{-10pt}
    \caption{}
    \label{fig:prob}
\end{subfigure}
\hspace{1pt}
\begin{subfigure}[t]{0.23\textwidth}
    \includegraphics[scale=0.48]{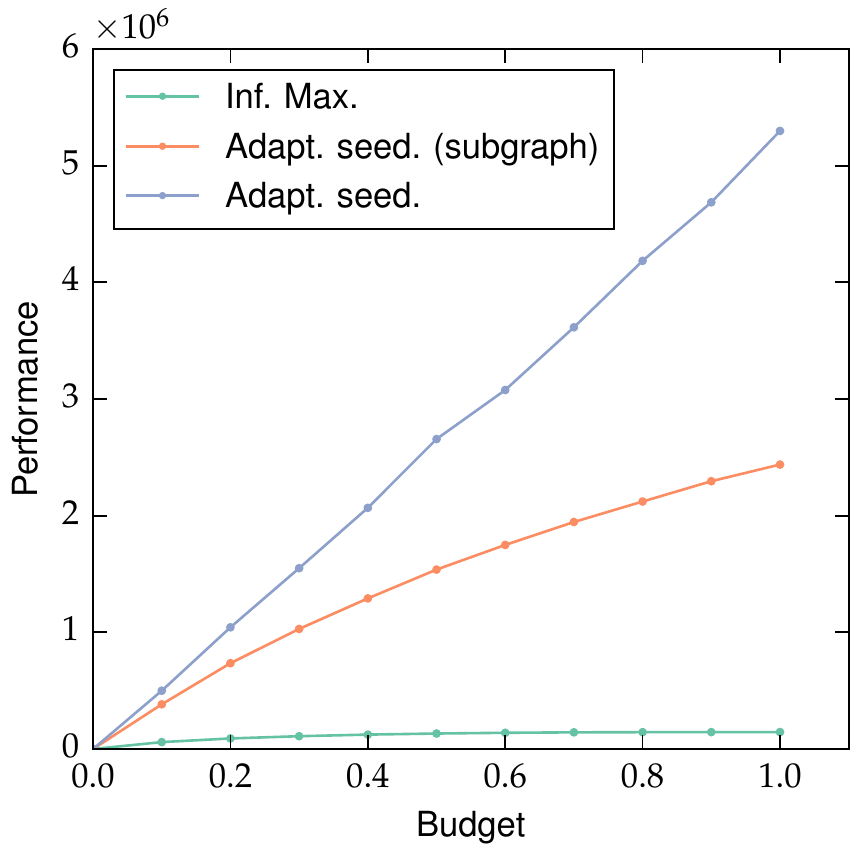}
    \caption{}
    \label{fig:killer}
     \end{subfigure}
 \vspace{-5pt}
 \caption{\small{(a) Performance of adaptive seeding for various propagation
     probabilities. (b) Performance of \emph{adaptive seeding} when restricted
 to the subgraph of users who \emph{liked} HBO (red line).}}
 \vspace{-20pt}
\end{figure}

\noindent\textbf{Impact of the Bernouilli parameter.} 
Figure~\ref{fig:prob} shows the impact of the probability of nodes realizing in
the second stage.  We computed the performance of \emph{adaptive seeding} when
each friend of a seeded user in the core set joins during the second stage
independently with probability $p$, using different values of $p$.  We call $p$
the \emph{Bernouilli} parameter, since the event that a given user joins on the
second stage of adaptive seeding is governed by a Bernouilli variable of
parameter $p$. We see that even with $p=0.01$, \emph{adaptive seeding} still
outperforms \textsf{IM}.  As $p$ increases, the performance of \emph{adaptive
seeding} quickly increases and reaches $80\%$ of the values of
Figure~\ref{fig:performance} at $p=0.5$.\newline

\noindent\textbf{Coarse estimation of probabilities.}  
In practice, the probability a user may be interested in promoting a campaign
her friend is promoting may vary. However, for those who have already expressed
interest in the promoted content, we can expect this probability to be close to
one.  We therefore conducted the following experiment.  We chose a page (HBO)
and trimmed the social graph we collected by only keeping on the second stage
users who indicated this page (HBO) in their list of interests.  This is
a coarse estimation of the probabilities as it assumes that if a friend follows
HBO she will be willing to promote with probability 1 (given a reward), and
otherwise the probability of her promoting anything for HBO is 0.
Figure~\ref{fig:killer} shows that even on this very restricted set of users,
\emph{adaptive seeding} still outperforms \textsf{IM} and reaches approximately
$50\%$ of the unrestricted adaptive seeding.\newline

\noindent\textbf{Impact of the probability distribution.} In order to test
scenarios where users have a rich spectrum of probabilities of realizing on the
second stage. We consider a setting where the Bernouilli parameter $p$ is drawn
from a distribution. We considered four different distributions; for each
distribution for fixed values of the budget and the parameter $p$, we tuned the
parameters of the distribution so that its mean is exactly $p$. We then plotted
the performance as a function of the budget and mean $p$. 

\begin{figure*}[t!]
\centering
\begin{subfigure}[b]{0.25\textwidth}
\includegraphics[width=\textwidth]{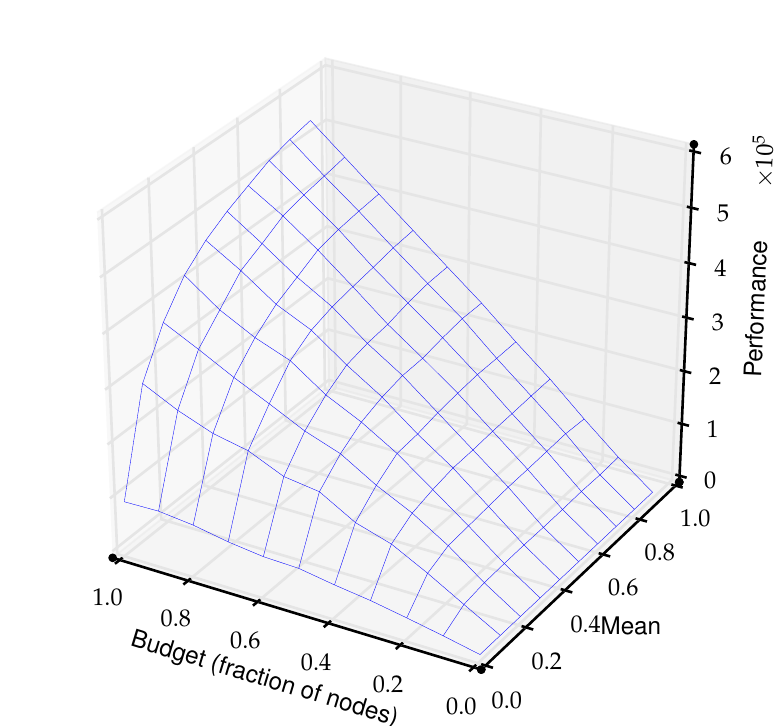}
\caption{Beta distribution}
\end{subfigure}
\begin{subfigure}[b]{0.25\textwidth}
\includegraphics[width=\textwidth]{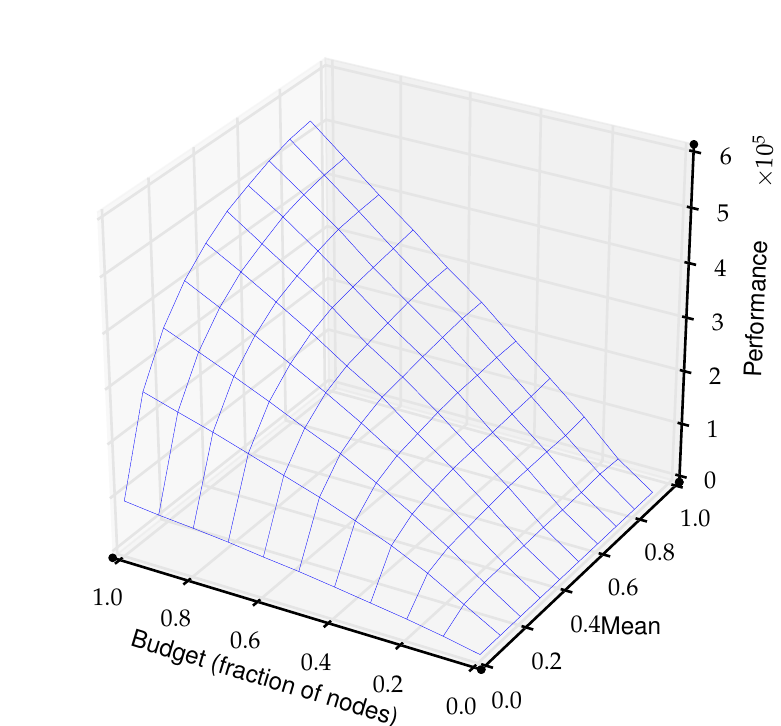}
\caption{Normal Distribution}
\end{subfigure}
\begin{subfigure}[b]{0.24\textwidth}
\includegraphics[width=\textwidth]{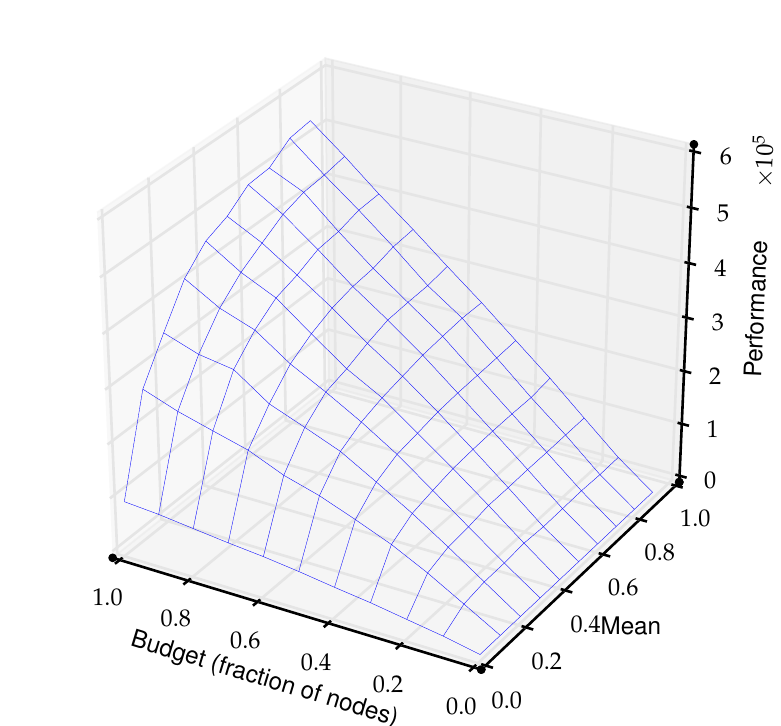}
\caption{Power-law distribution}
\end{subfigure}
\begin{subfigure}[b]{0.24\textwidth}
\includegraphics[width=\textwidth]{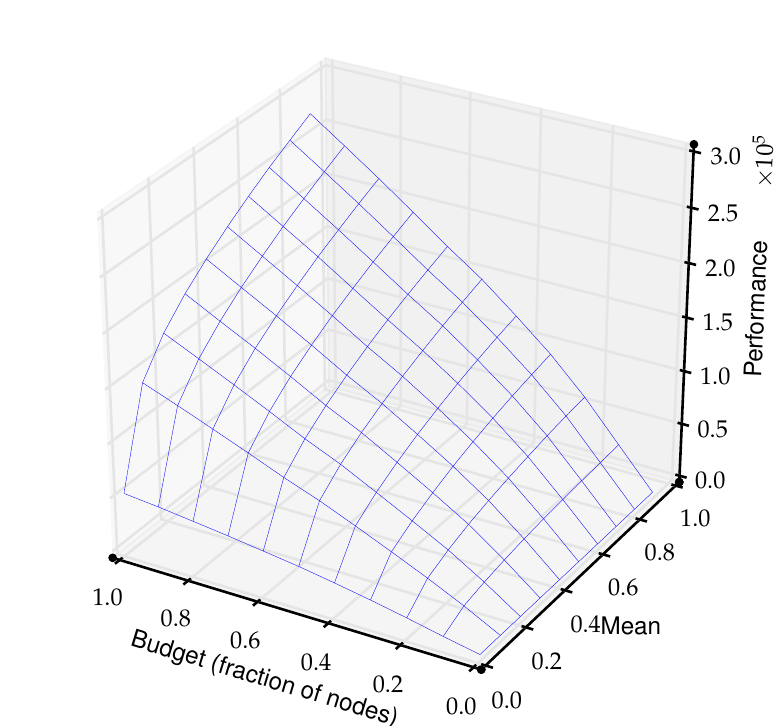}
\caption{Inverse degree}
\end{subfigure}
\caption{Performance of adaptive seeding as a function of the budget and the
mean of the distribution from which the Bernouilli parameters are drawn. The
details of the parameters for each distribution can be found in
Section~\ref{sec:robustness}.}.
\label{fig:bernouilli}
\vspace{-10pt}
\end{figure*}

For the Beta distribution, we fixed $\beta=5$ and tuned the $\alpha$ parameter
to obtain a mean of $p$, thus obtaining a unimodal distribution. For the normal
distribution, we chose a standard deviation of $0.01$ to obtain a distribution
more concentrated around its mean than the Beta distribution. Finally, for the
inverse degree distribution, we took the probability of a node joining on
the second stage to be proportional to the inverse of its degree (scaled so that on
average, nodes join with probability $p$).  The results are shown in
Figure~\ref{fig:bernouilli}.

We observe that the results are comparable to the one we obtained in the
uniform case in Figure~\ref{fig:prob} except in the case of the inverse degree
distribution for which the performance is roughly halved. Remember that the
value of a user $v$ on the second stage of adaptive seeding is given by $p_v
d_v$ where $d_v$ is its degree and $p_v$ is the its probability of realizing on
the second stage.  Choosing $p_v$ to be proportional to ${1}/{d_v}$ has the
effect of normalizing the nodes on the second stage and is a strong
perturbation of the original degree distribution of the nodes available on the
second stage.

\subsection{Impact of the Influence Model}
\label{sec:inf}

The Facebook Pages data set we collected is limited in that we only have access
to the 2-hop neighborhood around the seed users and we use the degree of the
second stage users as a proxy for their influence. As proved in
\cite{even-dar}, in the voter model, the influence of nodes converges to their
degree with high probability when the number of time steps become polynomially
large in the network size.

\begin{figure}
    \centering
    \includegraphics[width=0.48\textwidth]{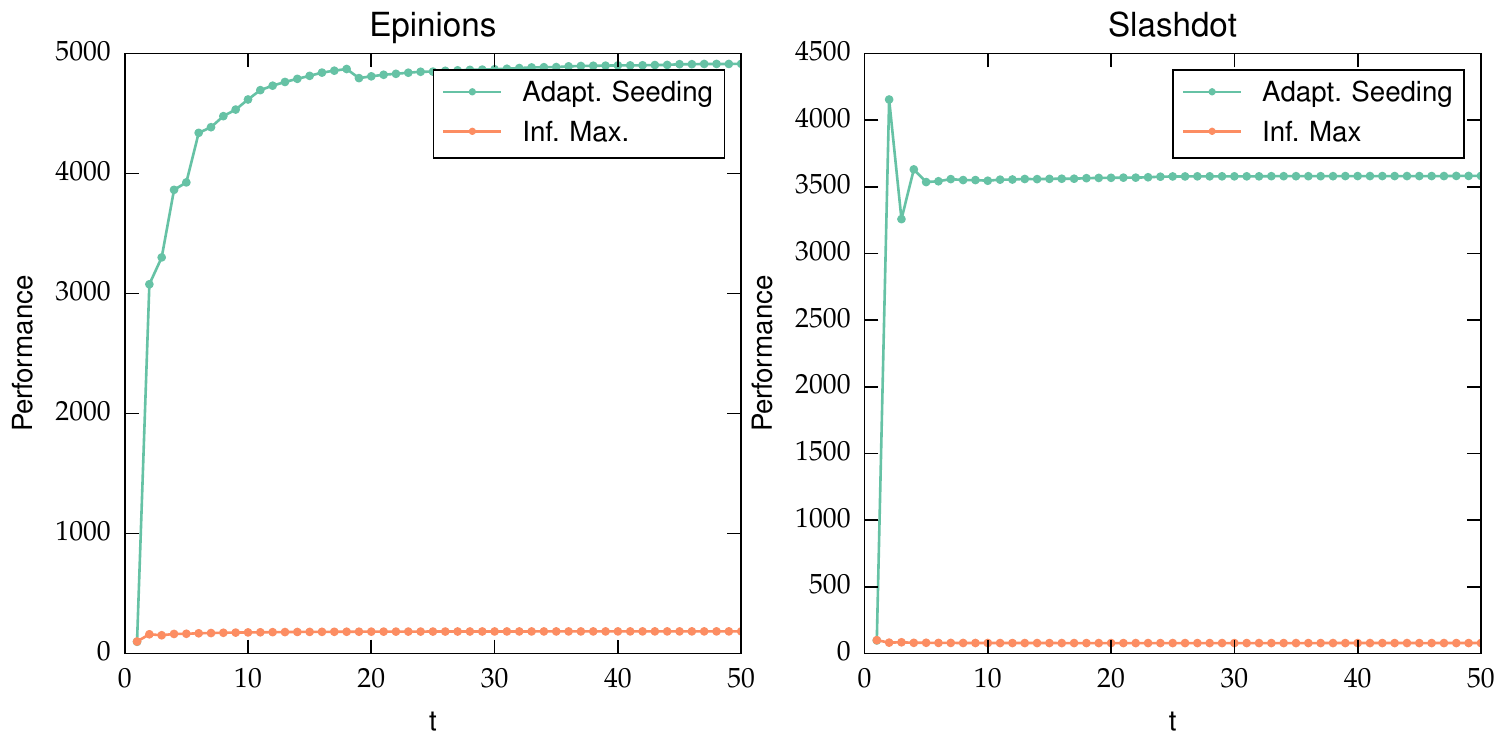}
    \vspace{-20pt}
    \caption{Performance of adaptive seeding compared to \textsf{IM} for the voter
    influence model with $t$ steps.}
    \vspace{-10pt}
    \label{fig:voter}
\end{figure}

In order to analyze the expected number of nodes influenced according to the
voter model that terminates after some fixed number of time steps, we use
publicly available data sets from \cite{snapnets} where the entire network is
at our disposal.  As discussed above, we sample nodes uniformly at random to
model the core set. We then run the voter model for $t$ time steps
to compute the influence of the second stage users. Figure~\ref{fig:voter}
shows the performance of adaptive seeding as a function of $t$ compared to the
performance of the \textsf{IM} benchmark. In this experiment, the budget was
set to half the size of the core set.

We see that the performance of adaptive seeding quickly converges (5 time steps
for \textsf{Slashdot}, 15 time steps for \textsf{Epinions}). In practice, the
voter model converges much faster than the theoretical guarantee of
\cite{even-dar}, which justifies using the degree of the second stage users as
measure of influence as we did for the Facebook Pages data sets.
Furthermore, we see that similarly to the Facebook data sets, adaptive seeding
significantly outperforms \textsf{IM}.

\subsection{Performance on Synthetic Networks} 

In order to analyze the impact of topological variations we generated synthetic
graphs using standard network models. All the generated graphs have $100,000$
vertices, for each model, we tuned the generative parameters to obtain when
possible a degree distribution (or graph density otherwise) similar to what we
observed in the Facebook Pages data sets.

\begin{figure}[t]
    \centering
    \includegraphics[width=0.48\textwidth]{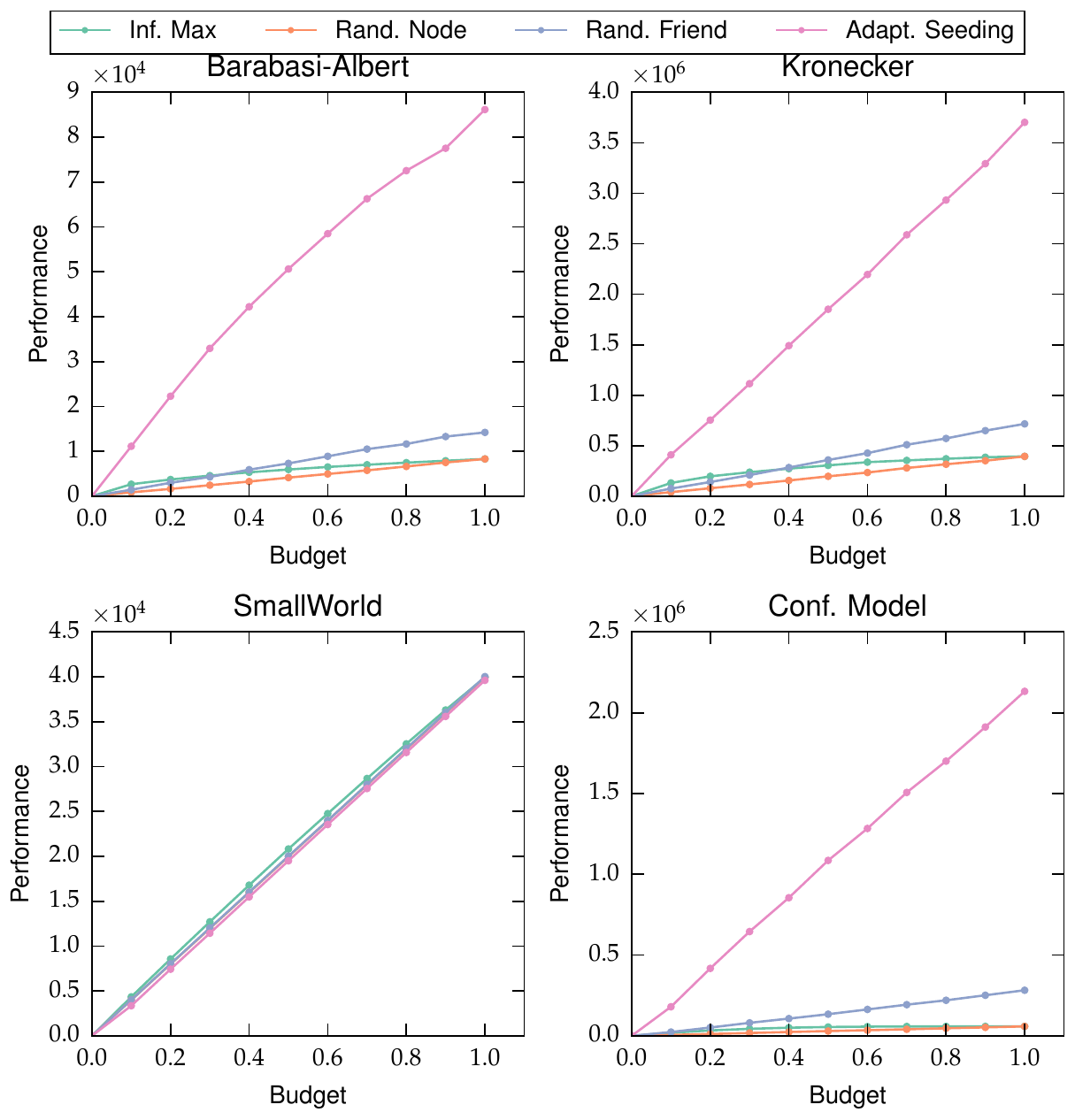}
    \vspace{-15pt}
    \caption{Performance of adaptive seeding on synthetic networks.}
    \label{fig:synth}
    \vspace{-10pt}
\end{figure}

\begin{itemize}
    \item \emph{Barabási-Albert:} this well-known model is often used to model
        social graphs because its degree distribution is a power law. We took
        10 initial vertices and added 10 vertices at each step, using the
        preferential attachment model, until we reached 100,000 vertices.
    \item \emph{Small-World:} also known as the Watts-Strogatz model. This
        model was one of the first models proposed for social networks. Its
        diameter and clustering coefficient are more representative of a social
        network than what one would get with the Erdős–Rényi model. We started
        from a regular lattice of degree 200 and rewired each edge with
        probability 0.3.
    \item \emph{Kronecker:} Kronecker graphs were more recently introduced in
        \cite{kronecker} as a scalable and easy-to-fit model for social
        networks. We started from a star graph with 4 vertices and computed
        Kronecker products until we reached 100,000 nodes.
    \item \emph{Configuration model:} The configuration model allows us to
        construct a graph with a given degree distribution. We chose a page
        (GAP) and generated a graph with the same degree distribution using the
        configuration model.
\end{itemize}
The performance of adaptive seeding compared to our benchmarks can be found in
Figure~\ref{fig:synth}. We note that the improvement obtained by adaptive
seeding is comparable to the one we had on real data except for the
\emph{Small-World} model. This is explained by the nature of the model:
starting from a regular lattice, some edges are re-wired at random. This model
has similar properties to a random graph where the friendship paradox does not
hold~\cite{LS13}. Since adaptive seeding is designed to leverage the friendship
paradox, such graphs are not amenable to this approach.

\subsection{Scalability}\label{sec:scalability}
To test the scalability of adaptive seeding we were guided by two central
questions.  First, we were interested to witness the benefit our non-sampling
approach has over the standard SAA method. Secondly, we wanted to understand
when one should prefer to use the LP-based approach from Section~\ref{sec:lp}
over the combinatorial one from Section~\ref{sec:comb}. The computations in
this section were run on Intel Core i5 CPU 4x2.40Ghz. For each computation, we
plot the time and number of CPU cycles it took.\newline

\noindent\textbf{Comparison with SAA.} The objective function of the
non-adaptive problem \eqref{eq:relaxed} is an expectation over exponentially
many sets, all possible realizations of the neighbors in the second stage.
Following the sampling-based approach introduced in \cite{singer}, this
expectation can be computed by averaging the values obtained in
$O\left(n^2\right)$ independent sample realizations of the second stage users
($n$ is the number of neighbors of core set users).  One important aspect of
the algorithms introduced in this paper is that in the additive case, this
expectation can be computed exactly without sampling, thus significantly
improving the theoretical complexity.

\begin{figure}[t]
    \centerline{ \includegraphics[width=0.48\textwidth]{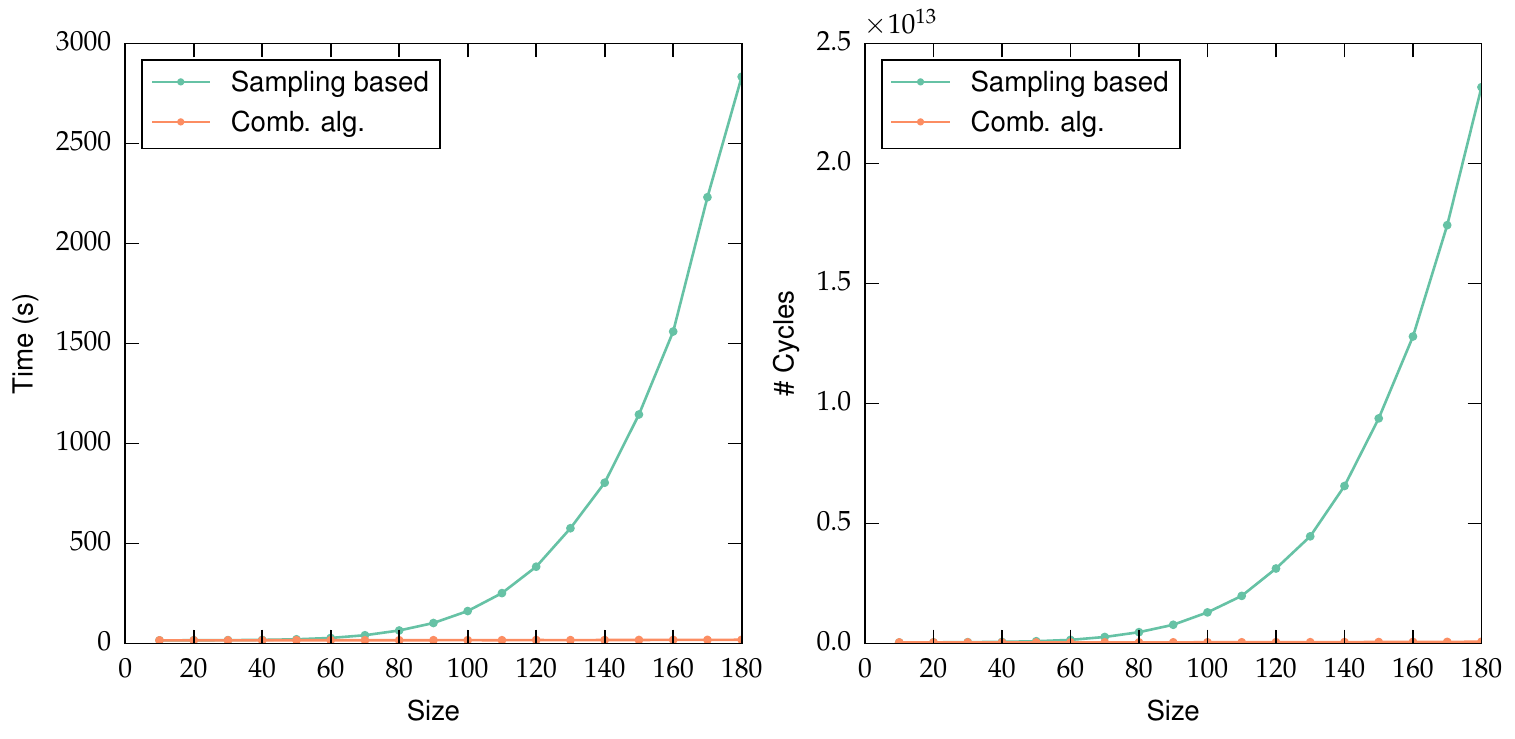} }
    \vspace{-10pt}
    \caption{Running time and number of CPU cycles used by the sampling-based
    algorithm and the combinatorial adaptive seeding algorithm for different
sizes of the core set.}
    \label{fig:sampling}
    \vspace{-10pt}
\end{figure}

In Figure~\ref{fig:sampling}, we compare the running time of our combinatorial
algorithm to the same algorithm where the expectation is computed via sampling.
We note that this sampling-based algorithm is still simpler than the algorithm
introduced in \cite{singer} for general influence models.  However, we observe
a significant gap between its running time and the one of the combinatorial
algorithm. Since each sample takes linear time to compute, this gap is in fact
$O(n^3)$, quickly leading to impracticable running times as the size of the
graph increases. This highlights the importance of the \emph{sans-sampling}
approach underlying the algorithms we introduced.\newline

\noindent\textbf{Combinatorial vs. LP algorithm.}
We now compare the running time of the LP-based approach and the combinatorial
approach for different instance sizes.

\begin{figure}[t]
    \centerline{ \includegraphics[scale=0.9]{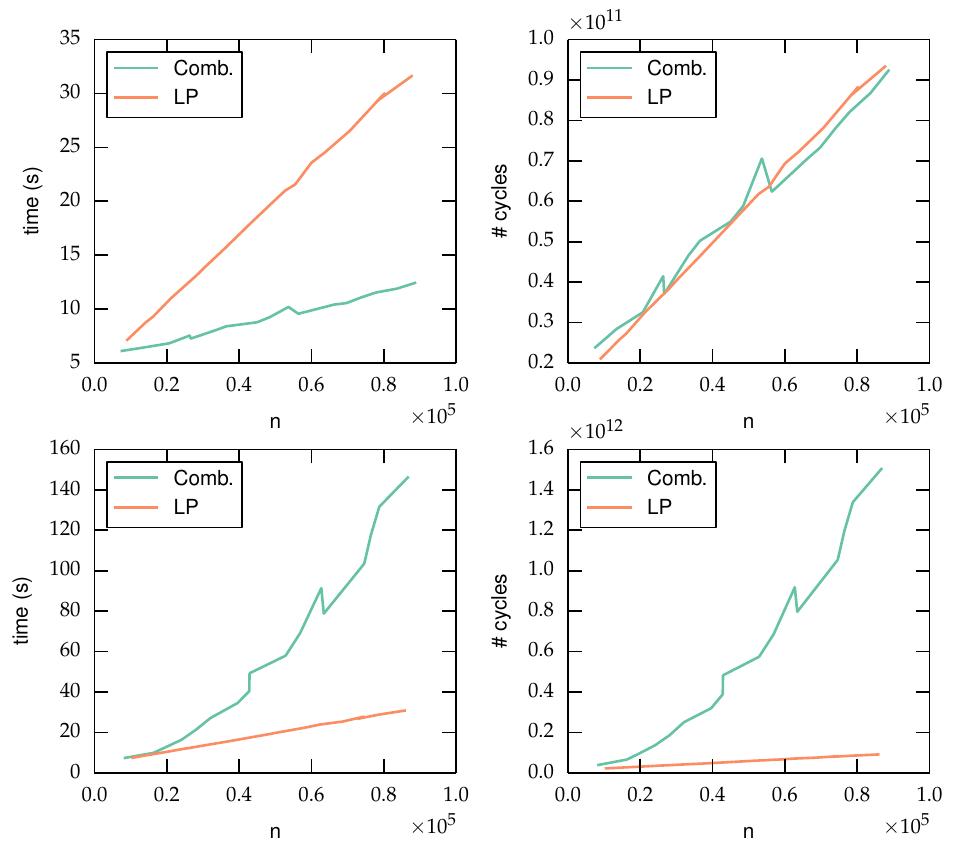} }
    \vspace{-10pt}
    \caption{Running time and number of CPU cycles of the combinatorial algorithm and the LP algorithm as a function of the number of nodes $n$. First row with budget $k=100$, second row with budget $k=500$.}
    \label{fig:time}
    \vspace{-10pt}
\end{figure}

Figure~\ref{fig:time} shows the running time and number of CPU cycles used by
the LP algorithm and the combinatorial algorithm as a function of the network
size $n$. The varying size of the network was obtained by randomly sampling
a varying fraction of core set users and then trimming the social graph by only
keeping friends of this random sample on the second stage.  The LP solver used
was CLP~\cite{clp}.

We observe that for a \emph{small} value of the budget $k$ (first row
of Figure~\ref{fig:time}), the combinatorial algorithm outperforms the LP
algorithm. When $k$ becomes \emph{large} (second row of
Figure~\ref{fig:time}), the LP algorithm becomes faster. This can be explained
by the $k^2$ factor in the running time of the combinatorial
algorithm (see Proposition~\ref{prop:running_time}). Even though the asymptotic
guarantee of the combinatorial algorithm should theoretically
outperform the LP-based approach for large $n$, we were not able to observe it
for our instance sizes. In practice, one can choose which of the two algorithms
to apply depending on the relative sizes of $k$ and $n$.\newline

%\begin{figure}
%    \centerline{ \includegraphics[width=0.47\textwidth]{images/sampling.pdf} }
%    \vspace{-10pt}
%    \caption{}
%    \label{fig:sampling}
%\end{figure}

%\noindent \textbf{Others}
%\begin{itemize}
%\item One plot with Twitter data on the same verticals/ same posts.
%\item Run pipeline on available data sets, with random samples;
%\item It's true that adaptive seeding is looking at more people, but so what?  If you waited a day, it would still be better: take a vertical, split it at random to two days.  Perform adaptive seeding on one day, and compare to influence max. on two days.
%\item compare to other influence maximization algorithms?
%\item test on several public datasets
%\item LP vs combinatorial algorithm vs sampling running time
%\end{itemize}

\section{Related work}
Influence maximization was introduced by Domingos and Richardson~\cite{DR01, RD02}, formulated by Kempe, Kleinberg and Tardos~\cite{KKT03, KKT05}, and has been extensively studied since~\cite{MR07, C08, LKGFVG07,MR07,C08,KDD11,borgs2012influence}.  
The main result in~\cite{KKT03, KKT05} is a characterization of influence processes as submodular functions, which implies good approximation guarantees for the influence maximization problem. 
%Kempe, Kleinberg and Tardos~\cite{KKT03, KKT05} were able to cast this problem as a submodular optimization problem for many classes of influence models, hence allowing for algorithms with good approximations guarantees. 
%Further refinements of the influence models and associated approximation guarantees can be found in \cite{MR07, C08}.  
In~\cite{even-dar}, the authors look at the special case of the voter model and design efficient algorithms in this setting.

Our two-stage model for influence maximization is related to the field of stochastic optimization where problems are commonly solved using the \emph{sample average approximation} method~\cite{SampleAverage}. %In \cite{dean2004approximating, gupta2012approximation}, the authors use a different notion of non-adaptive solutions.
%The problem we study is the same as in \cite{singer} but our goals are different. 
%Our main goal here is to obtain scalable algorithms that can be applied on the datasets we collected.
%Other models of adaptive optimization have been previously studied in the context of influence maximization.  
%In \cite{asadpour2008stochastic}, the authors study a stochastic sequential submodular maximization problem where at each step an element is chosen, its realization is revealed and the next decision is made. 
%Golovin and Krause \cite{golovin2011adaptive} study a similar model and apply it to a multi-stage influence maximization problem. We note that contrary to our model, the decision made at a given stage does not affect the following stages as the entire set of nodes is available as potential seeds at every stage.
Golovin and Krause \cite{golovin2011adaptive} study a stochastic sequential submodular maximization problem where at each step an element is chosen, its realization is revealed and the next decision is made. We note that contrary to adaptive seeding, the decision made at a given stage does not affect the following stages as the entire set of nodes is available as potential seeds at every stage.

\section*{Acknowledgement}
This research is supported in part by a Google Research Grant and NSF grant CCF-1301976.

\bibliographystyle{abbrv}
%\balancecolumns % GM June 2007
\balance
\bibliography{main}

\appendix
\section{Adaptivity proofs}
\label{sec:ad-proofs}
\begin{proof}[of Proposition~\ref{prop:gap}]
    We will first show that the optimal adaptive policy can be interpreted as
    a non-adaptive policy. Let $S$ be the optimal adaptive
    solution and define $\delta_R:\neigh{X}\mapsto \{0,1\}$:
    \begin{displaymath}
        \delta_R(u) \defeq \begin{cases}
            1&\text{if } u\in\argmax\big\{f(T);\; T\subseteq R,\; |T|\leq
        k-|S|\big\} \\
            0&\text{otherwise}
        \end{cases},
    \end{displaymath}
    one can write
    \begin{displaymath}
    \begin{split}
        \sum_{R\subseteq\neigh{S}} p_R
        \max_{\substack{T\subseteq R\\|T|\leq k-|S|}}f(T)
        &=
        \sum_{R\subseteq\neigh{S}} p_R
        \sum_{u\in\neigh{X}}\delta_R(u)w_u\\
        &=
        \sum_{u\in\neigh{X}}w_u\sum_{R\subseteq\neigh{S}}p_R\delta_R(u).
    \end{split}
    \end{displaymath}

    Let us now define for $u\in\neigh{X}$:
    \begin{displaymath}
        q_u \defeq \begin{cases}
            \sum_{R\subseteq\neigh{S}}\frac{p_R}{p_u}\delta_R(u)
            &\text{if }p_u\neq 0\\
            0&\text{otherwise}
        \end{cases}.
    \end{displaymath}
    This allows us to write:
    \begin{displaymath}
            \sum_{R\subseteq\neigh{S}} p_R
            \max_{\substack{T\subseteq R\\|T|\leq k-|S|}}f(T)
            = \sum_{u\in\neigh{X}}p_uq_uw_u = F(\mathbf{p}\circ\mathbf{q})
    \end{displaymath}
    where the last equality is obtained from \eqref{eq:multi} by successively using the linearity of the expectation and the linearity of $f$.

    Furthermore, observe that $q_u\in[0,1]$, $q_u=0$ if $u\notin\neigh{S}$ and:
    \begin{displaymath}
        \begin{split}
        |S|+\sum_{u\in\neigh{X}}p_uq_u
        &= |S|+\sum_{R\subseteq\neigh{S}}p_R\sum_{u\in\neigh{X}}\delta_R(u)\\
        &\leq |S| + \sum_{R\subseteq\neigh{S}}p_R(k-|S|)\leq k
    \end{split}
    \end{displaymath}

    Hence, $(S,\mathbf{q})\in\mathcal{F}_{NA}$. In other words, we have written the optimal adaptive solution as a relaxed
    non-adaptive solution. This conclude the proof of the proposition.
\end{proof}

\vspace{0.5em}

\begin{proof}[of Proposition~\ref{prop:cr}]
    Using the definition of $\mathrm{A}(S)$, one can write:
    \begin{displaymath}
        \mathrm{A}(S) = \sum_{R\subseteq\neigh{S}} p_R
        \max_{\substack{T\subseteq R\\|T|\leq k-|S|}}f(T)
        \geq \sum_{R\subseteq\neigh{S}} p_R \mathbf{E}\big[f(I)\big]
    \end{displaymath}
    where the inequality comes from the fact that $I$ is a feasible random set: $|I|\leq k-|S|$, hence the expected value of $f(I)$ is bounded by the maximum of $f$ over feasible sets.

Equation~\eqref{eq:cr} then implies:
\begin{equation}\label{eq:tmp}
    \mathrm{A}(S) 
    \geq (1-\varepsilon)\sum_{R\subseteq\neigh{S}} p_R F(\mathbf{q})
    = (1-\varepsilon)F(\mathbf{p}\circ\mathbf{q}).
\end{equation}

Equation~\eqref{eq:tmp} holds for any $\varepsilon\geq 0$. In particular, for $\varepsilon$ smaller than $\inf_{S\neq T} |A(S)-A(T)|$, we obtain that $\mathrm{A}(S)\geq F(\mathbf{p}\circ\mathbf{q})$. Note that such a $\varepsilon$ is at most polynomially small in the size of the instance.
$(S, \mathbf{q})$ is an $\alpha$-approximate non adaptive solution, hence $F(\mathbf{p}\circ\mathbf{q}) \geq \alpha\mathrm{OPT}_{NA}$. We can then conclude by applying Proposition~\ref{prop:gap}. 
\end{proof}

\section{Algorithms Proofs}
\label{sec:alg-proofs}
We first discuss the NP-hardness of the problem.

\noindent \textbf{\textsc{NP}-Hardness.} In contrast to standard influence maximization, adaptive seeding is already \textsc{NP}-Hard even for the simplest cases.  In the case when $f(S)=|S|$ and all probabilities equal one, the decision problem is whether given a budget $k$ and target value $\ell$ there exists a subset of $X$ of size $k-t$ which yields a solution with expected value of $\ell$ using $t$ nodes in $\mathcal{N}(X)$.  This is equivalent to deciding whether there are $k-t$ nodes in $X$ that have $t$ neighbors in $\mathcal{N}(X)$.  To see this is \textsc{NP}-hard, consider reducing from \textsc{Set-Cover} where there is one node $i$ for each input set $T_i$, $1\leq i\leq n$, with $\neigh{i}= T_i$ and integers $k,\ell$, and the output is ``yes'' if there is a family of $k$ sets in the input which cover at least $\ell$ elements, and ``no'' otherwise.

\subsection{LP-based approach}
\label{sec:lp-proofs}
In the LP-based approach we rounded the solution using the pipage rounding method.  We discuss this with greater detail here.

\noindent \textbf{Pipage Rounding.}
The pipage rounding method~\cite{pipage} is a deterministic rounding method that can be applied to a variety of problems.  In particular, it can be applied to LP-relaxations of the \textsc{Max-K-Cover} problem where we are given a family of sets that cover elements of a universe and the goal is to find $k$ subsets whose union has the maximal cardinality.  The LP-relaxation is a fractional solution over subsets, and the pipage rounding procedure then rounds the allocation in linear time, and the integral solution is guaranteed to be within a factor of $(1-1/e)$ of the fractional solution. 
We make the following key observation: for any given $\textbf{q}$, one can remove all elements in $\mathcal{N}(X)$ for which $q_{u}=0$, without changing the value of any solution $(\boldsymbol\lambda,\textbf{q})$.
Our rounding procedure can therefore be described as follows: given a solution $(\boldsymbol\lambda,\textbf{q})$ we remove all nodes $u \in \mathcal{N}(X)$ for which $q_{u}=0$, which leaves us with a fractional solution to a (weighted) version of the \textsc{Max-K-Cover} problem where nodes in $X$ are the sets and the universe is the set of weighted nodes in $\mathcal{N}(X)$ that were not removed.  We can therefore apply pipage rounding and lose only a factor of $(1-1/e)$ in quality of the solution.

\subsection{Combinatorial Algorithm}
\label{sec:comb-proofs}
We include the missing proofs from the combinatorial algorithm section.  The scalability and implementation in MapReduce are discussed in this section as well.

\begin{proof}[of Lemma~\ref{lemma:nd}]
\emph{W.l.o.g} we can rename and order the pairs in $T$ so that $w_1\geq w_2\geq\ldots\geq w_m $.
Then, $\mathcal{O}(T,b)$ has the following simple piecewise linear expression:
\begin{displaymath}\label{eq:pw}
    \mathcal{O}(T,b) = 
    \begin{cases}
        b w_1&\text{if }0\leq b<p_1\\
        \displaystyle
        \sum_{k=1}^{i-1}p_k(w_k-w_i)
        + b w_i
        &\displaystyle\text{if } 0\leq b - \sum_{k=1}^{i-1}p_k< p_i\\
        \displaystyle
        \sum_{k=1}^m p_kw_k
        &\displaystyle\text{if } b\geq\sum_{i=1}^m p_k

    \end{cases}
\end{displaymath}

Let us define for $t\in\reals^+$, $n(t)\defeq \inf\Big\{i\text{ s.t.
} \sum_{k=1}^i p_k > t\Big \}$ with $n(t)=+\infty$ when the set is empty. In
particular, note that $x\mapsto n(t)$ is non-decreasing. Denoting
$\partial_+\mathcal{O}_T$ the right derivative of $\mathcal{O}(T,\cdot)$, one
can write $\partial_+\mathcal{O}_T(t)=w_{n(t)}$, with the convention that
$w_\infty = 0$.

Writing $i \defeq \sup\Big\{j\text{ s.t.
} w_j\geq w_x\Big\}$, it is easy to see that
$\partial_+\mathcal{O}_{T\cup\{x\}}\geq\partial_+\mathcal{O}_T$. Indeed:
\begin{enumerate}
    \item if $n(t)\leq i$ then $\partial_+\mathcal{O}_{T\cup\{x\}}(t)
        = \partial_+\mathcal{O}_T(t)= w_{n(t)}$.
    \item if $n(t)\geq i+1$ and $n(t-c)\leq i$ then $\partial_+\mathcal{O}_{T\cup\{x\}}(t)
        = w_x\geq w_{n(t)}= \partial_+\mathcal{O}_T(t)$.
    \item if $n(t-c)\geq i+1$, then $\partial_+\mathcal{O}_{T\cup\{x\}}
        = w_{n(t-c)}\geq w_{n(t)}=\partial_+\mathcal{O}_T(t)$.
\end{enumerate}

Let us now consider $b$ and $c$ such that $b\leq c$. Then, using the integral
representation of $\mathcal{O}(T\cup\{x\},\cdot)$ and $\mathcal{O}(T,\cdot)$, we get:
\begin{multline*}
    \mathcal{O}(T\cup\{x\},c)-\mathcal{O}(T\cup\{x\},b)=\int_b^c\partial_+\mathcal{O}_{T\cup\{x\}}(t)dt\\
    \geq\int_b^c\partial_+\mathcal{O}_T(t)dt=\mathcal{O}(T,c)-\mathcal{O}(T,b)
\end{multline*}
Re-ordering the terms, $\mathcal{O}(T\cup\{x\},c)-\mathcal{O}(T,c)\geq
\mathcal{O}(T\cup\{x\},b)-\mathcal{O}(T,b)$
which concludes the proof of the lemma.
\end{proof}

\vspace{0.5em}

\begin{proof}[of Lemma~\ref{lemma:sub}]
    Let $T\subseteq\neigh{X}$ and $x, y\in\neigh{X}\setminus T$. Using the
    second-order characterization of submodular functions, it suffices to show
    that:
    \begin{displaymath}\label{eq:so}
        \mathcal{O}(T\cup\{x\},b)-\mathcal{O}(T,b)\geq
        \mathcal{O}(T\cup\{y, x\},b)-\mathcal{O}(T\cup\{y\},b)
    \end{displaymath}

    We distinguish two cases based on the relative position of $w_x$ and $w_y$.
    The following notations will be useful: $S_T^x \defeq \big\{u\in
    T\text{ s.t.  }w_x\leq w_u\big\}$ and $P_T^x\defeq
    T\setminus S_T^x$.

    \textbf{Case 1:} If $w_y\geq w_x$, then one can
    write:
    \begin{gather*}
        \mathcal{O}(T\cup\{y,x\},b) = \mathcal{O}(P_T^y\cup\{y\},b_1)+
        \mathcal{O}(S_T^y\cup\{x\},b_2)\\
        \mathcal{O}(T\cup\{y\},b) = \mathcal{O}(P_T^y\cup\{y\},b_1)
        + \mathcal{O}(S_T^y,b_2)
    \end{gather*}
    where $b_1$ is the fraction of the budget $b$ spent on $P_T^y\cup\{y\}$ and
    $b_2=b-b_1$.
    
    Similarly:
    \begin{gather*}
        \mathcal{O}(T\cup\{x\},b) = \mathcal{O}(P_T^y, c_1) + \mathcal{O}(S_T^y\cup\{x\},c_2)\\
        \mathcal{O}(T, b) = \mathcal{O}(P_T^y, c_1) + \mathcal{O}(S_T^y,c_2)
    \end{gather*}
    where $c_1$ is the fraction of the budget $b$ spent on $P_T^y$ and $c_2
    = b - c_1$. 

    Note that $b_1\geq c_1$: an optimal solution will first spent as much
    budget as possible on $P_T^y\cup\{y\}$ before adding elements in
    $S_T^y\cup\{x\}$.

    In this case:
    \begin{displaymath}
        \begin{split}
            \mathcal{O}(T\cup\{x\},b)-\mathcal{O}(T,b)&=
        \mathcal{O}(S_T^y\cup\{x\},c_2)+\mathcal{O}(S_T^y,c_2)\\
        &\geq \mathcal{O}(S_T^y\cup\{x\},b_2)+\mathcal{O}(S_T^y,b_2)\\
        & = \mathcal{O}(T\cup\{y, x\},b)-\mathcal{O}(T\cup\{y\},b)
    \end{split}
    \end{displaymath}
    where the inequality comes from Lemma~\ref{lemma:nd} and 
    $c_2\geq b_2$.

    \textbf{Case 2:} If $w_x > w_y$, we now decompose
    the solution on $P_T^x$ and $S_T^x$:
    \begin{gather*}
        \mathcal{O}(T\cup\{x\},b) = \mathcal{O}(P_T^x\cup\{x\},b_1)
        + \mathcal{O}(S_T^x,b_2)\\
        \mathcal{O}(T,b) = \mathcal{O}(P_T^x,c_1)+\mathcal{O}(S_T^x,c_2)\\
        %\intertext{and}
        \mathcal{O}(T\cup\{y, x\},b) = \mathcal{O}(P_T^x\cup\{x\},b_1)
        + \mathcal{O}(S_T^x\cup\{y\},b_2)\\
        \mathcal{O}(T\cup\{y\},b) = \mathcal{O}(P_T^x,c_1)+\mathcal{O}(S_T^x\cup\{y\},c_2)
    \end{gather*}
    with $b_1+b_2=b$, $c_1+c_2=b$ and $b_2\leq c_2$. 

    In this case again:
        \begin{multline*}
            \mathcal{O}(T\cup\{x\},b)-\mathcal{O}(T,b)=
        \mathcal{O}(S_T^x,b_2)-\mathcal{O}(S_T^x,c_2)\\
        \geq \mathcal{O}(S_T^x\cup\{y\},b_2)-\mathcal{O}(S_T^x\cup\{y\},c_2)\\
        = \mathcal{O}(T\cup\{y, x\},b)-\mathcal{O}(T\cup\{y\},b)
    \end{multline*}
    where the inequality uses Lemma~\ref{lemma:nd} and $c_2\geq b_2$.

    In both cases, we were able to obtain the second-order characterization of submodularity. This concludes the proof of the lemma.
\end{proof}

\vspace{0.5em}

\begin{proof}[of Proposition~\ref{prop:sub}]
    Let us consider $S$ and $T$ such that $S\subseteq T\subseteq X$ and $x\in
    X\setminus T$. In particular, note that $\neigh{S}\subseteq\neigh{T}$. 

    Let us write $\neigh{S\cup\{x\}}=\neigh{S}\cup R$ with $\neigh{S}\cap
    R=\emptyset$ and similarly, $\neigh{T\cup\{x\}}=\neigh{T}\cup R'$ with
    $\neigh{T}\cap R'=\emptyset$. It is clear that $R'\subseteq R$. Writing $R'=\{u_1,\ldots,u_k\}$:
    \begin{multline*}
            \mathcal{O}(\neigh{T\cup\{x\}},b)- \mathcal{O}(\neigh{T},b)\\
            =\sum_{i=1}^k\mathcal{O}(\neigh{T}\cup\{u_1,\ldots u_i\},b)
            -\mathcal{O}(\neigh{T}\cup\{u_1,\ldots u_{i-1}\},b)\\
            \leq \sum_{i=1}^k\mathcal{O}(\neigh{S}\cup\{u_1,\ldots u_i\},b)
            -\mathcal{O}(\neigh{S}\cup\{u_1,\ldots u_{i-1}\},b)\\
            =\mathcal{O}(\neigh{S}\cup R',b)-\mathcal{O}(\neigh{S},b)
    \end{multline*}
    where the inequality comes from the submodularity of $\mathcal{O}(\cdot,b)$ proved in Lemma~\ref{lemma:sub}. This same function is also obviously set-increasing, hence:
    \begin{multline*}
            \mathcal{O}(\neigh{S}\cup R',b)-\mathcal{O}(\neigh{S},b)\\
            \leq \mathcal{O}(\neigh{S}\cup R,b)-\mathcal{O}(\neigh{S},b)\\
            =\mathcal{O}(\neigh{S\cup\{x\}},b)-\mathcal{O}(\neigh{S},b)
    \end{multline*}
    This concludes the proof of the proposition.
\end{proof}

\begin{proof}[of Proposition~\ref{prop:main_result}]
    We simply note that the content of the outer \textsf{for} loop on line 2 of Algorithm~\ref{alg:comb} is the greedy submodular maximization algorithm of \cite{nemhauser}. Since $\mathcal{O}(\neigh{\cdot}, t)$ is submodular (Proposition~\ref{prop:sub}), this solves the inner $\max$ in \eqref{eq:sub-mod} with an approximation ratio of $(1-1/e)$. The outer \textsf{for} loop then computes the outer $\max$ of \eqref{eq:sub-mod}.

    As a consequence, Algorithm~\ref{alg:comb} computes a $(1-1/e)$-approximate non-adaptive solution. We conclude by applying Proposition~\ref{prop:cr}.
\end{proof}

\subsection{Parallelization}
\label{sec:para}
As discussed in the body of the paper, the algorithm can be parallelized across $k$ different machines, each one computing an approximation for a fixed budget $k-t$ in the first stage and $t$ in the second.
A slightly more sophisticated approach is to consider only $\log n$ splits: $(1,k-1),(2,k-2),\ldots,(2^{\lfloor \log n \rfloor},1)$ and then select the best solution from this set.  It is not hard to see that in comparison to the previous approach, this would reduce the approximation guarantee by a factor of at most 2: if the optimal solution is obtained by spending $t$ on the first stage and $k-t$ in the second stage, then since $t \leq 2\cdot2^{\lfloor \log t \rfloor}$ the solution computed for $(2^{\lfloor \log t \rfloor}, k - 2^{\lfloor \log t \rfloor})$ will have at least half that value.  
More generally, for any $\epsilon>0$ one can parallelize over $\log_{1+\epsilon}n$ machines at the cost of losing a factor of $(1+\epsilon)$ in the approximation guarantee.

\subsection{Implementation in MapReduce}
\label{sec:mr}

As noted in Section~\ref{sec:comb}, lines 4 to 7 of Algorithm~\ref{alg:comb}
correspond to the greedy heuristic of \cite{nemhauser} applied to the
submodular function $f_t(S) \defeq \mathcal{O}\big(\neigh{S}, t\big)$.
A variant of this heuristic, namely the $\varepsilon$-greedy heuristic,
combined with the \textsc{Sample\&Prune} method of \cite{mr} allows us to write
a MapReduce version of Algorithm~\ref{alg:comb}. The resulting algorithm is
described in Algorithm~\ref{alg:combmr}

\begin{algorithm}
    \caption{Combinatorial algorithm, MapReduce}
    \label{alg:combmr}
    \algsetup{indent=2em}
    \begin{algorithmic}[1]
        \STATE $S\leftarrow \emptyset$
        \FOR{$t=1$ \TO $k-1$}
            \STATE $S_t\leftarrow \emptyset$
            \FOR{$i=1$ \TO $\log_{1+\varepsilon}\Delta$}
                \STATE $U\leftarrow X$, $S'\leftarrow \emptyset$
                \WHILE{$|U|>0$}
                    \STATE $R\leftarrow$ sample from $U$ w.p. $\min\left(1,
                    \frac{\ell}{|U|}\right)$
                    \WHILE{$|R|>0$ \OR $|S_t\cup S'|< k$}
                        \STATE $x\leftarrow$ some element from $R$
                        \IF{$\nabla f_t(S_t\cup S', x)\geq\frac{\Delta}{(1+\varepsilon)^i}$}
                            \STATE $S'\leftarrow S'\cup\{x\}$
                        \ENDIF
                        \STATE $R\leftarrow R\setminus\{x\}$
                    \ENDWHILE
                    \STATE $S_t\leftarrow S_t\cup S'$
                    \STATE $U\leftarrow\{x\in U\,|\, \nabla f_t(S_t,
                    x)\geq\frac{\Delta}{(1+\varepsilon)^i}\}$
                \ENDWHILE
            \ENDFOR
            \IF{$\mathcal{O}(\neigh{S_t},t)>\mathcal{O}(\neigh{S},k-|S|)$}
                \STATE $S\leftarrow S_t$
            \ENDIF
        \ENDFOR
        \RETURN $S$
    \end{algorithmic}
\end{algorithm}

We denoted by $\nabla f_t(S, x)$ the marginal increment of $x$ to the set $S$
for the function $f_t$, $\nabla f_t(S, x) =  f_t(S\cup\{x\}) - f_t(S)$.
$\Delta$ is an upper bound on the marginal contribution of any element. In our
case, $\Delta = \max_{u\in\neigh{X}} w_u$ provides such an upper bound. The
sampling in line 7 selects a small enough number of elements that the
\texttt{while} loop from lines 8 to 14 can be executed on a single machine.
Furthermore, lines 7 and 16 can be implemented in one round of MapReduce each.

The approximation ratio of Algorithm~\ref{alg:combmr} is
$1-\frac{1}{e}-\varepsilon$. The proof of this result as well as the optimal
choice of $\ell$ follow from Theorem 10 in \cite{mr}.

\end{document}